\documentclass[manuscript]{geophysics}
\usepackage{latexsym, amsthm, amssymb, graphicx, float}
\usepackage{natbib}
\usepackage{bm}
\newcommand  {\x}{{\bf x}}
\newcommand  {\m}{{\bf m}}
\newcommand  {\bd}{{\bf d}}
\newcommand  {\rr}{{\bf r}}
\newcommand  {\z}{{\bf z}}
\newcommand  {\dd}{{\bf d}}

\newcommand  {\y}{{\bf y}}
\newcommand  {\LL}{{\bf L}}

\usepackage{xcolor}
\makeatletter

\begin{document}

\title{ {Deep Convolutional Neural Network and Sparse Least Squares Migration}}
\author{Zhaolun Liu\footnotemark[1], Yuqing Chen\footnotemark[2] and Gerard Schuster\footnotemark[3]}
\address{\footnotemark[1] Formerly  King Abdullah University of Science and Technology, Department of Earth Science and Engineering, Thuwal, Saudi Arabia; presently Princeton University, Department of Geosciences,  Princeton, NJ 08544, USA.  E-mail: zhaolunl@princeton.edu\\
	   \footnotemark[2] Formerly  King Abdullah University of Science and Technology, Department of Earth Science and Engineering, Thuwal, Saudi Arabia; presently CSIRO, Deep Earth Imaging Future Science Platform, Kensington, Australia; E-mail: Yu.Chen@csiro.au\\
	   \footnotemark[3] King Abdullah University of Science and Technology, 
	   Department of Earth Science and Engineering, Thuwal, Saudi Arabia, 23955-6900.
	   Email: gerard.schuster@kaust.edu.sa.
	 }
\lefthead{Liu et al}
\righthead{\emph{CNN and NNLSM}}
\maketitle

\begin{abstract}
We recast the forward pass of a multilayered convolutional neural network (CNN) as the solution 
to the problem of sparse least squares migration (LSM). The CNN filters and feature maps are shown 
to be analogous, but not equivalent, to the migration Green's functions and the quasi-reflectivity distribution,
respectively.
This provides a physical interpretation of the filters and feature
maps in deep CNN in terms of the operators for seismic imaging.
Motivated by the connection between sparse LSM
and CNN, we propose the neural network version of
sparse LSM. Unlike the standard LSM method that finds the
optimal reflectivity image, neural network LSM (NNLSM) finds both the optimal quasi-reflectivity
image and the quasi-migration Green's functions. These quasi-migration-Green's functions are also denoted
as the convolutional filters in a CNN and are similar to migration Green's functions.
The advantage of NNLSM over standard LSM is that its computational cost is significantly less and
it can be used for denoising coherent and incoherent noise in migration images.
Its disadvantage is that the NNLSM quasi-reflectivity image is
only an approximation to the actual reflectivity distribution.
However, the quasi-reflectivity image can be used as a superresolution attribute image
for high-resolution delineation of geologic bodies.

\end{abstract}

\section{Introduction}

Deep convolutional neural networks (CNNs) have been recently used for solving 
geophysical problems, such as
seismic first-arrival picking \citep{Kai2018picking, Yuan2018,Hu2019},
seismic interpretation \citep{Wu2019gji,Wu2019FaultNet3DPF,Shi2019salt}, and seismic imaging and inversion 
\citep{Xu2019PINN,Sun2020RNNFWI,Kaur2020}. 
Interpretation examples include salt classification \citep{waldeland2018convolutional,Shi2019salt},
fault detection \citep{huang2017scalable,xiong2018seismic,Wu2019FaultNet3DPF,zheng2019applications},
reservoir characterization \citep{karimpouli2010new,cao2017time,zhu2017inversion},  and 
seismic lithofacies classification \citep{ross2017comparison,liu2019seismic}
with semi-supervised learning \citep{di2020seismic}.
Other applications  include the use of  neural networks for well-log interpolation 
\citep{saggaf2003estimation,salehi2017estimation,pham2020missing}, 
seismic data interpolation \citep{mandelli2018seismic,mandelli2019interpolation,wang2020seismic},
velocity model building \citep{araya2018deep,richardson2018seismic},
well-log ties \citep{bader2019missing}, synthetic well-data generation \citep{rolon2009using},
autoencoders for  unsupervised facies analysis \citep{qian2018unsupervised}, and supervised horizon tracking
\citep{peters2019multiresolution,peters2019neural}.
Recently, an unsupervised autoencoder method with regularization
was developed by \citet{shi2020waveform} to track target horizons.

There are many types of neural network
or machine learning methods, selections ranging from
generative adversarial networks for seismic interpolation \citep{siahkoohi2018seismic},
residual networks for traces missing at regular intervals \citep{wang2019deep},
Monte Carlo and support vector regression \citep{jia2018intelligent} for data interpolation,
autoencoders \citep{wang2020seismic,shi2020waveform} for target horizons tracking,
and recurrent neural networks for well-log interpolation \citep{pham2020missing}.
The design of geophysical CNN architectures  has largely been based on 
empirical evidence from computer vision research, insights from the principles of 
artificial intelligence  and heuristic experimentation. Heuristic
experimentation is most often used to decide the parameter design\footnote{The design
of a CNN architecture selects the number
of CNN layers, the number of filters/layer, the size of the filters,
the type of activation functions, the number of skip layers and
whether a layer acts as a decoding or encoding operation.}
for the CNN architecture, which has both merits and liabilities.
The merit is that trial-and-error with different architecture parameters
is likely to give excellent results
for a particular data set, but it might not be the best one for a completely different data set.
This shortcoming in using empirical tests for parameter selection
largely results from the absence of a rigorous mathematical foundation
\citep{papyan2016convolutional,papyan2017convolutional} for neural networks in general, and CNN in particular.

To mitigate this problem for CNN-based imaging algorithms, we now present a physical interpretation of  the CNN filters  and feature maps in terms of the physics-based operators for seismic imaging. With such an understanding, we can use a physics-based rationale for the better design of CNN architectures in seismic migration and inversion.

\citet{Donoho2019} points out that
``machine learning has a troubled relationship with understanding the foundation
of its achievements well and its literature is admittedly corrupted by
anti-intellectual and anti-scholarly tendencies''.
Progress in advancing the capabilities of deep neural networks will be severely
stymied unless its mathematical foundations are established.
As a first step in this direction,
\citet{papyan2016convolutional} proposed
that the forward modeling operation of CNN could be recast as finding the sparsest
model $\m$ under the $L_1$ norm subject to honoring
the data misfit constraint $||{{\boldsymbol \Gamma} \m-\bd}||_2^2 \leq \beta$:
\begin{eqnarray}
Given:&& {\boldsymbol \Gamma}, \bd~~~~\mbox{and}~~{{\boldsymbol \Gamma} \m } = {\bd} + \boldsymbol{noise},\nonumber\\
Find: &&{\m^*}=\arg \min_{\m} ||\m||_1 ~~~~\nonumber\\
&&\mbox{subject~to}~||{{\boldsymbol \Gamma} \m -\bd }||_2^2 \leq \beta\,
\label{MLL.eq1}
\end{eqnarray}
where $\m^*$ is the optimal solution of $\m$, ${\boldsymbol \Gamma}$ is 
the dictionary matrix, $\bd$ represents the signal, and the scalar $\beta$ 
is the specified noise tolerance.
The iterative solution to this problem is a series
 of forward-modeling operations of
a neural network, where the mathematical operations of each layer consist
two steps: a  weighted summation of input values to give the vector $\z$ followed by
a two-sided soft thresholding operation denoted as $\sigma( {\z})$ \citep{papyan2016convolutional}.

The sparse constraint problem defined in equation~\ref{MLL.eq1} 
is commonly seen in geophysics, for example, the least square migration (LSM) with a sparse constraint.
LSM is an important  seismic-imaging technique to produce images with better balanced amplitudes,
fewer artifacts and better resolution than standard migration \citep{lailly1983seismic,tarantola1987inverse,schuster1993least,nemeth1999least,chavent1999optimal,duquet2000kirchhoff,feng2017elastic,schuster2019least}.
The sparse constraint is one
of the important regularization terms used in solving the ill-conditioned least-squares problem  
\citep{sacchi1995high, de2002sparse, kuhl2003least,wang2005sparse} and
sparse LSM (SLSM) has been demonstrated to be effective for
mitigation of aliasing artifacts and crosstalk in LSM \citep{wang2007high,herrmann2009curvelet,
  dutta2017sparse, witte2017sparsity,li2018sparse}. The image-domain 
  SLSM finds the sparsest reflectivity $\m^*$ with sparse constraints 
  to minimize the objective function
$||{{\boldsymbol \Gamma} \m-\m^{mig}}||_2^2$, where $\boldsymbol \Gamma$ is the Hessian 
matrix and $\m^{mig}$ is the migration image.
Here, we can see SLSM shares the same problem as that defined in equation ~\ref{MLL.eq1}.
Following the work of \citet{papyan2016convolutional},
we show that the sparse solution to the LSM problem 
reduces to the forward modeling operations of a multilayered neural network.
The CNN filters and feature maps are shown 
to be analogous, but not equivalent, to the migration Green's functions (Hessian) and the 
reflectivity distribution.

The standard SLSM algorithm needs to solve the wave equation, which is time-consuming.
Motivated by the connection between sparse LSM
and CNN, we propose the neural network version of
sparse LSM, which does not need to solve the wave equation and is faster than standard SLSM.
Instead of just finding the optimal reflectivity $\m^*$,
we optimize for both the quasi-reflectivity $\bf m$
and the quasi-migration-Green's functions $\boldsymbol \Gamma$.
These quasi-migration-Green's functions approximate the role of migration
Green's function \citep{schuster2000green} and are denoted
as the convolutional filters in a convolutional neural network.
As discussed in Appendix \ref{A}, the migration Green's function is the point-scatterer 
response of the migration operator.
The final image is denoted as the neural network  least squares migration (NNLSM) estimate of
the quasi-reflectivity distribution that honors the $L_1$  sparsity condition.
The next section shows the connection between the
 multilayer neural network and the solution to the multilayer NNLSM problem.
 This is followed by
 the numerical examples with the synthetic models and field
data from the North Sea.

\section{Theory of Neural Network Least Squares Migration}
The theory of standard image-domain LSM is first presented
to establish the benchmark solution where the optimal reflectivity function
minimizes the image misfit under the $L_2$ norm. This is then
followed by the derivation of the sparse least squares migration (SLSM)
solution for a single-layer network.
The final two subsections derive the NNLSM solution for  single-layer 
and multilayer networks, respectively.

\subsection{Least Squares Migration}

The least squares migration (LSM)  problem can be
defined \citep{schuster2000green,schuster2017seismic} as
finding the reflectivity coefficients $m_i$
in the $N \times 1$ vector $\m$
that minimize the $L_2$ objective function $\epsilon=1/2||{\boldsymbol \Gamma}{\bf m}-{\bf m}^{mig}||_2^2$,
\begin{eqnarray}
\m^* &=&
\arg \min_{\m} \{ \frac{1}{2} ||{\boldsymbol \Gamma}\m - \m^{mig} ||_2^2\} ,
\label{SLSM.eq1a}
\end{eqnarray}
where  ${\boldsymbol \Gamma}=\LL^T \LL$ is the symmetric $N\times N$ Hessian
matrix, $\LL$ is the forward modeling operator, and $\LL^T$ is the migration
operator. Here, $\m^{mig} = {\LL^T} \dd$ is the  migration image computed by
migrating the recorded data $\dd$ with the migration operator $\LL^T$.
Alternatively, the  image-domain  LSM problem can also be defined as
finding $\bf m$ that minimizes $\epsilon=1/2({\bf m}^T {\boldsymbol \Gamma} {\bf m}-{\bf m}^T{\bf m}^{mig})$,
which has a more well-conditioned solution than the one in equation 2 \citep{schuster2017seismic}. However, we will use
equation \ref{SLSM.eq1a} as the definition of the LSM problem in order to be consistent
with the notation from \citet{papyan2017convolutional}. 
The kernel associated with the Hessian matrix $\LL^T \LL$ is also
known as the point scatterer
response of the migration operator or the migration Green's function \citep{schuster2000green}.
It is a square matrix that is assumed  to be invertible,
otherwise a regularization term is incorporated into the objective function.

A formal solution to equation~\ref{SLSM.eq1a} is
\begin{eqnarray}
\m^*&=& {\boldsymbol \Gamma }^{-1} \m^{mig},
\end{eqnarray}
where it is too expensive to directly
compute the inverse Hessian ${\boldsymbol \Gamma }^{-1}$.
Instead, a gradient method
gives the iterative solution
\begin{eqnarray}
\m^{(k+1)}&=& \m^{(k)}-\alpha  {\boldsymbol \Gamma}^T ({\boldsymbol \Gamma} \m^{(k)}-\m^{mig}),
\end{eqnarray}
where $\alpha$ is the step length, $\boldsymbol \Gamma$ is symmetric,
and $\m^{(k)}$ is the
solution at the $k^{th}$ iteration.
Typically, a regularization term is used to stabilize the solution, for example, the sparse constraint which
will be introduced in the next subsection.

\subsection{Sparse Least Squares Migration}

The sparse least squares migration (SLSM) in the image domain is defined as
finding the reflectivity coefficients $m_i$
in the $N \times 1$ vector $\m$
that minimize the objective function $\epsilon$ \citep{perez2013estimating}:
\begin{eqnarray}
\epsilon &=&
\frac{1}{2} ||{\boldsymbol \Gamma}\m - \m^{mig} ||_2^2 + \lambda S(\m),
\label{SLSM.eq1}
\end{eqnarray}
where ${\boldsymbol \Gamma}=\LL^T \LL$
represents the migration
Green's function \citep{schuster2000green},
 $\lambda>0$ is a positive scalar, $\m^{mig} = {\LL^T} \dd$ is the  migration image,
and $S(\m)$ is a sparseness function. For example,
the sparseness function might be $S(\m)=||\m||_1$ or
$S(\m)=log(1+||\m||_2^2)$.

The solution
to equation~\ref{SLSM.eq1} is
\begin{eqnarray}
\m^*&=&\arg\min_{\m} {\Big [} \frac{1}{2} ||{\boldsymbol \Gamma}\m - \m^{mig} ||_2^2 + \lambda S(\m) {\Big ]},
\label{SLSM.eq1aa}
\end{eqnarray}
which can be approximated by an iterative
gradient descent method:
\begin{eqnarray}
m_i^{(k+1)} &=& m_i^{(k)} - \alpha {\Big [}{\boldsymbol \Gamma}^T\overbrace{({\boldsymbol \Gamma}\m - \m^{mig})}^{\rr=residual} + \lambda S(\m)'{\Big ]}_i,\nonumber\\
&=& m_i^{(k)} - \alpha [{\boldsymbol \Gamma}^T \rr + \lambda S(\m)']_i.
\label{SLSM.eq2}
\end{eqnarray}
Here, $S(\m)_i^{\prime}$ is the derivative of the sparseness function with respect
to the model parameter $m_i$
and the step length is $\alpha$.
Vectors and matrices  are denoted by boldface lowercase
and uppercase letters, respectively.  When  $S(\m)=||\m||_1$,
the iterative solution in
equation~\ref{SLSM.eq2} can be recast as
\begin{eqnarray}
 m_i^{(k+1)} =  {\mbox {soft}}({\Big [}\m^{(k)} -\frac{1}{\alpha} {\boldsymbol \Gamma}^T({\boldsymbol \Gamma} \m^{(k)}-\m_{mig}){\Big ]}_i,\frac{\lambda}{\alpha} ),
\label{SLSM.aeq2}
\end{eqnarray}
where, {\it soft}  is the two-sided soft thresholding function \citep{papyan2016convolutional} 
derived in Appendix \ref{B} (see equation~\ref{SLSM.Aeq4}).
Here, $\boldsymbol \Gamma={\bf L}^T{\bf L} $ is computed by solving the wave equation to get the
forward modeled field and backpropagating the data by a numerical solution to the adjoint wave equation.

\begin{figure}
  \centering
\includegraphics[width=0.9\columnwidth]{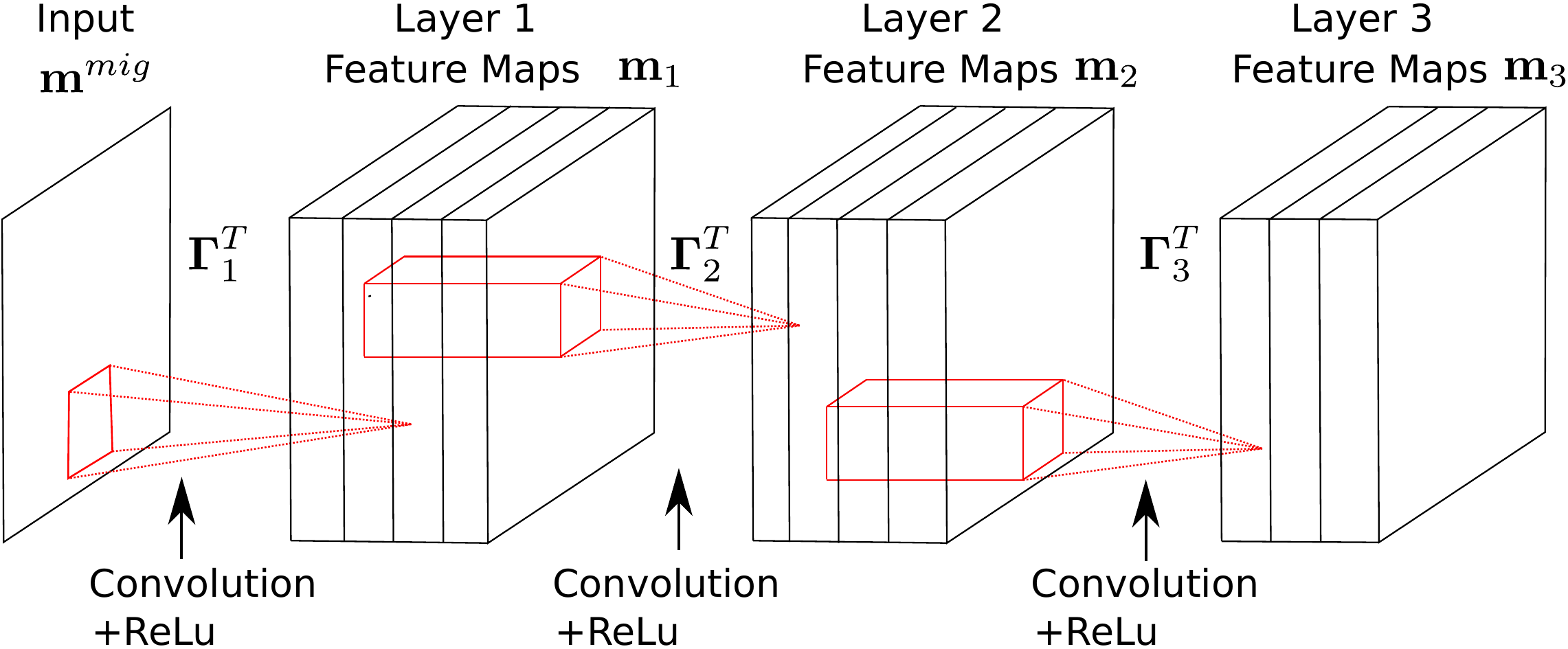}
\caption{The forward modeling procedure for a multilayer CNN is equivalent to the multilayer sparse solution.  }
\label{zhaolun.fig1_0}
\end{figure}

Equation~\ref{SLSM.aeq2} is similar to the forward modeling operation associated with  the first layer of
the neural network in Figure~\ref{zhaolun.fig1_0}.
That is, set  $k=0$,
$\m^{(0)}=0$, $\alpha=1$, and let the
input vector be the scaled residual vector $\rr=-({\boldsymbol \Gamma} \m^{(0)}-\m_{mig})=\m_{mig}$ so that
the first-iterate solution can be compactly represented by
\begin{eqnarray}
\m^{(1)}=\mbox{soft}({\boldsymbol \Gamma}^T \m^{mig},\lambda).
\label{ch.sparse.eq1}
\end{eqnarray}
Here, the input vector $\rr=\m^{mig}$ is
multiplied by the matrix ${\boldsymbol \Gamma}^T$ to give $\z={\boldsymbol \Gamma}^T \rr$,
and the elements of = $\z$ are then thresholded and shrunk to give the output
${\m}=\mbox{soft}(\z,\lambda)$. If we impose a positivity constraint for $\z$
and a shrinkage constraint so $\lambda$ is small,
then the soft thresholding function becomes that of a one-sided threshold function, also known as the Rectified Linear Unit or ReLU function.
To simplify the notation, the $\mbox{soft}(\z,\lambda)$ function or $\mbox{ReLu}(\z)$ function
is replaced by $\sigma_{\lambda}(\z)$ so that
equation~\ref{ch.sparse.eq1} is given by
\begin{eqnarray}
\m^{(1)}=\sigma_{\lambda} ({\boldsymbol \Gamma}^T \m^{mig}).
\label{ch.sparse.eq1aa}
\end{eqnarray}
For the ReLu function there is no shrinkage so $\lambda=0$. However, $\boldsymbol \Gamma$ in equation \ref{ch.sparse.eq1aa}
is computed by forward and backward solutions to the wave equation. Unlike a neural network,
the physics of wave propagation is included with the sparse LSM solution in equation \ref{SLSM.aeq2}.

\subsection{Neural Network Least Squares Migration}

We now propose the neural network version of SLSM
that finds both  $\boldsymbol \Gamma^*$ and ${\bf m}^*$ which minimize
equation~\ref{SLSM.eq1}, which is equivalent to the convolutional sparse coding (CSC)
problem. We denote the optimal solution $\bf m$ as  the neural network least squares
migration (NNLSM) image.
Here, we assume that the migration
image $\m^{mig}$ can be decomposed into components that have the form
${\boldsymbol \Gamma}_1\m_1$,
where $\m_1$ represents a sparse quasi-reflectivity structure for the
$1^{th}$ CNN layer in Figure \ref{zhaolun.fig1_0} and
${\boldsymbol \Gamma}_1$ has a convolutional structure.
The solution can be found by using the Alternating Direction Method of
Multipliers (ADMM) method either in the Fourier domain \citep{heide2015fast} or
in the space domain \citep{papyan2017_ICCV}, which alternates between
finding $\boldsymbol \Gamma^*$ (dictionary learning problem)
and then finding ${\bf m}^*$ (sparse pursuit problem).
The key difference between least squares sparse inversion and NNLSM is
that $\boldsymbol \Gamma$ is not computed by numerical solutions to the wave equation.
Instead, the coefficients of $\boldsymbol \Gamma$ are
computed by the usual gradient-descent learning algorithm of
a convolutional neural network. For this reason,
we denote $\bf m^*$ as the quasi-reflectivity distribution and
$\boldsymbol \Gamma^*$ as the quasi-migration Green's function.

Appendix \ref{C}
shows the general solution for NNLSM for a single-layer neural network,
where the optimal $\boldsymbol \Gamma^*$ is composed of the quasi-migration Green's functions,
which are denoted as convolutional filters
in the machine learning terminology \citep{liu2018neural, liu2019csc}.
Each filter is used to compute a feature map that corresponds to
a sub-image of quasi-reflection coefficients in the context of LSM.

We now compute the NNLSM image for a 1D model, where we assume
 $\m^{mig}$ is a N-dimensional vector which can be expressed as,
\begin{equation}
  \m^{mig}=\sum_i^{k_0} {\bm\gamma}_i*\m_{1i}'.
  \label{eqn:1dcsc}
\end{equation}
Here, ${\bm\gamma}_i $ is the $i^{th}$
local filter with length of $n_0$, $\m'_{1i}$ is
the $i^{th}$ feature map, ``*'' denotes the
convolution operator and $k_0$ is the number of the filters.
Alternatively, following Figure~\ref{fig:1dmatrixform}a,
equation~\ref{eqn:1dcsc} can be written in matrix form as
$\m^{mig}={\boldsymbol \Gamma}_1\m_1={\boldsymbol \Gamma}_1'\m_1'$ \citep{papyan2017convolutional},
where ${\boldsymbol \Gamma}_1$ is a convolutional
matrix containing in its columns the $k_0$ filters with all of their shifts.
${\boldsymbol \Gamma}_1'$ is a concatenation of banded and circulant
\footnote{We shall assume throughout this paper that boundaries are
  treated by a periodic continuation, which gives rise to the cyclic
structure.} matrices, which is the same as ${\boldsymbol \Gamma}_1$
except that the order of the columns is different.
$\m_1'$ is a concatenation of the feature map vectors $\m_{1i}'$ for $i=1,2,\cdots,k_0$.

The advantage of NNLSM is that only inexpensive
matrix-vector multiplications are used and  no expensive solutions to the
wave equation are needed for backward and forward
wavefield propagation. As will be seen later, convolutional filters
that appear to be coherent noise can be excluded for denoising the migration image.

\begin{figure}[htpb]
  \centering
  \includegraphics[width=\columnwidth]{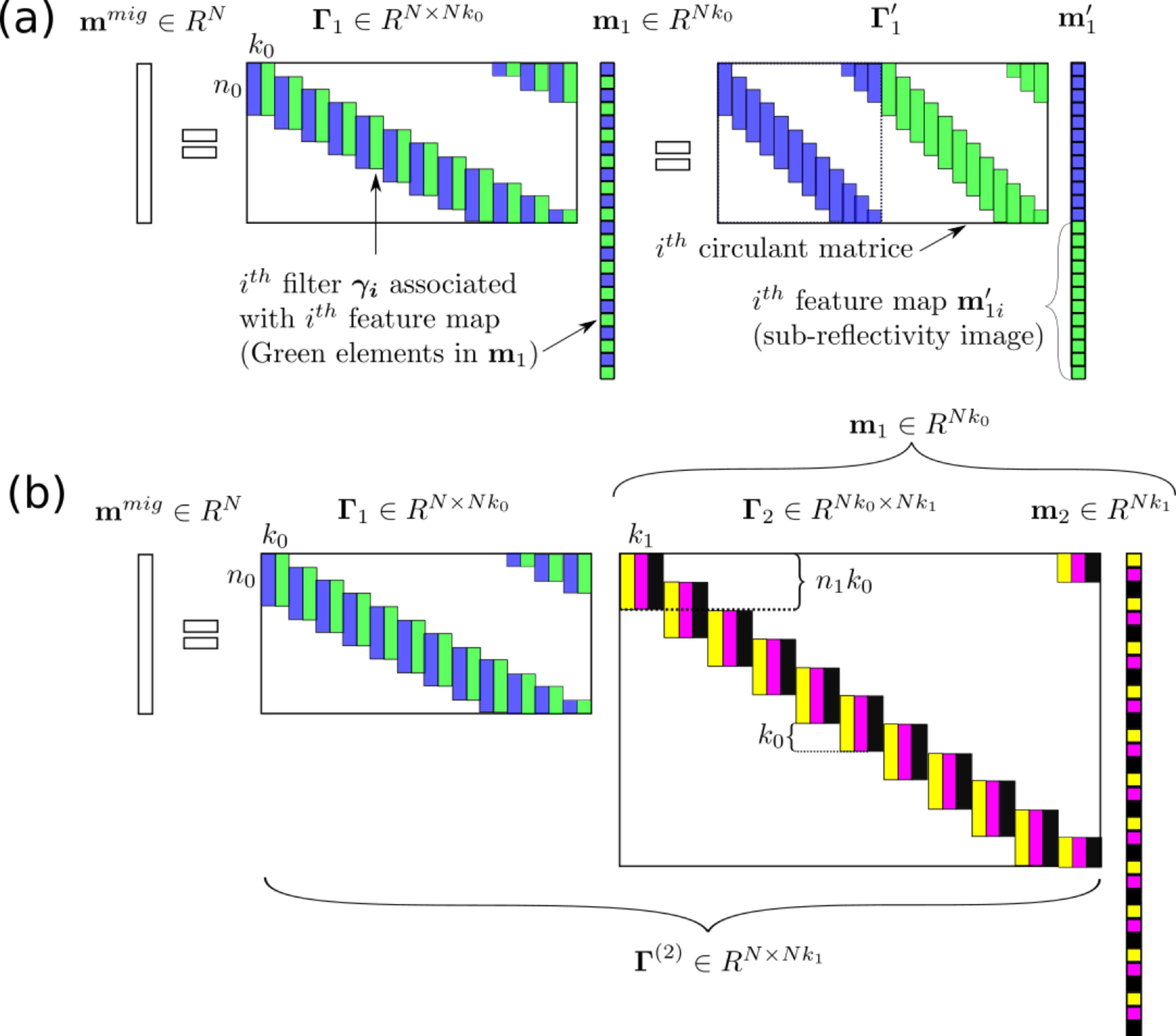}
  \caption {(a) Single-layer NNLSM and (b) multilayer NNLSM for a one-dimensional migration image $\m^{mig}$.}
  \label{fig:1dmatrixform}
\end{figure}

\subsection{Multilayer Neural Network  LSM}

The multilayer NNLSM is a natural extension of the single-layer NNLSM.  For NNLSM,
the migration image $\m^{mig}$ can be expressed as
$\m^{mig}={\boldsymbol \Gamma}_1\m_1$
(Figure~\ref{fig:1dmatrixform}a), where there are
$k_0$ filters in ${\boldsymbol \Gamma}_1$ and $k_0$ sub-quasi-reflectivity images
in $\m_1$. Following \citet{sulam2018multilayer}, we can cascade this
model by imposing a similar
assumption to the sparse representation
$\m_1$,  i.e., $\m_1={\boldsymbol \Gamma}_2\m_2$, for
a corresponding convolutional matrix
${\boldsymbol \Gamma}_2$ with $k_1$ local
filters and a sparse sub-quasi-reflectivity image $\m_2$,
as depicted in Figure~\ref{fig:1dmatrixform}b. In this case,
the filter size is $n_1\times k_0$ and there are $k_1$ sub-quasi-reflectivity
images in $\m_2$.

Similar to the derivation by \citet{papyan2017convolutional} and \citet{sulam2018multilayer}, the
multilayer neural network LSM problem is defined as
the following.
\begin{eqnarray}
\mbox{Find:} &&\m_i, \Gamma_i~~~\mbox{such~that}\nonumber\\
\m_1^*&=&\arg\min_{\m_1,\Gamma_1} {\Big [} \frac{1}{2} ||{\boldsymbol \Gamma}_1\m_1 - \m^{mig} ||_2^2 + \lambda S(\m_1) {\Big ]},\nonumber\\
\m_2^*&=&\arg\min_{\m_2,\Gamma_2} {\Big [} \frac{1}{2} ||{\boldsymbol \Gamma}_2\m_2 - \m_1^* ||_2^2 + \lambda S(\m_2) {\Big ]},\nonumber\\
&\vdots&\nonumber\\
\m_{N}^*&=&\arg\min_{\m_N,\Gamma_N} {\Big [} \frac{1}{2} ||{\boldsymbol \Gamma}_N\m_N - \m_{N-1}^* ||_2^2 + \lambda S(\m_N) {\Big ]},
\label{eqn:multilayerSLSM}
\end{eqnarray}
where ${\boldsymbol \Gamma}_i$ is the $ith$ Hessian matrix in the $ith$ layer.
The first iterate solution to the above  system of equations can be cast in
a form similar to equation~\ref{ch.sparse.eq1aa}, except we have
\begin{eqnarray}
\m_N^*&\approx & \sigma_{\lambda}~({\boldsymbol \Gamma}_N^T ~~\sigma_{\lambda}({\boldsymbol \Gamma}_{N-1}^T(...\sigma_{\lambda}({\boldsymbol \Gamma}_1^T \m^{mig})...),
\end{eqnarray}
which is a repeated concatenation of the two
operations of a multilayered neural network: matrix-vector multiplication followed by
a thresholding operation.  In all cases, we use a convolutional neural network where
different filters are applied to the input from the previous layer
to give  feature maps associated with the next layer, as shown in Figure~\ref{zhaolun.fig1_0}.

For a perfect prediction of the migration image, $\m^{mig}$ can also be approximated
as $\m^{mig}={\boldsymbol \Gamma}_1{\boldsymbol \Gamma}_2\dots{\boldsymbol \Gamma}_N\m_N$. We refer to ${\boldsymbol \Gamma}^{(i)}$ as
the effective filter at the $i^{th}$ level,
\begin{equation}
{\boldsymbol \Gamma}^{(i)}={\boldsymbol \Gamma}_1{\boldsymbol \Gamma}_2\dots{\boldsymbol \Gamma}_i,
  \label{eqn:effectiveFilter}
\end{equation}
so that
\begin{equation}
  \m^{mig}={\boldsymbol \Gamma}^{(i)}\m_i.
\end{equation}
The next section tests the effectiveness of NNLSM on both synthetic data and field data.

\section{Numerical Results}

We now present numerical simulations of NNLSM.
Instead of only determining the optimal reflectivity
$\bf m$ as computed by SLSM, the NNLSM method computes
both quasi-reflectivity $\bf m$ and the elements of the Hessian matrix
${\boldsymbol \Gamma}=\bf L^T{\bf L}$.
Each block of ${\boldsymbol \Gamma}$ is considered to be
the {\it segment response function} (SSF)
of the migration operator rather than the {\it point spread function} (PSF).
If the actual Green's functions are used to construct ${\boldsymbol \Gamma}$
then each column of the Hessian matrix is the point scatterer
response of the migration operator \citep{schuster2000green}.
In contrast, the NNLSM  Hessian  is composed
of blocks, where each block is the segment scatterer response of
the migration operator. An example will be shown later where a segment
of the reflector is migrated to give the migration segment
response of the migration operator. The computational cost for computing SSF's is several
orders of magnitude less than that for PSFs because no solutions to the wave equation are needed.
The penalty, however, is that the resulting solution $\bf m$ is 
not the true reflectivity,
but a sparse representation of it we denote as the quasi-reflectivity distribution.

Using the terminology of neural networks, we can also denote
the sparse sub-quasi-reflectivity images as feature maps. Each block in $ \boldsymbol \Gamma$ will
be denoted as a filter.
Therefore the vector output of ${\boldsymbol \Gamma} \bf m$ can be
interpreted as a sum of filter vectors $ {\bm \gamma}_i$ weighted by the
coefficients in $\bf m$, where ${\bm \gamma}_i$ is
the $i^{th}$ column vector of ${\boldsymbol \Gamma}$.

\begin{figure}
  \centering
  \includegraphics[width=\columnwidth]{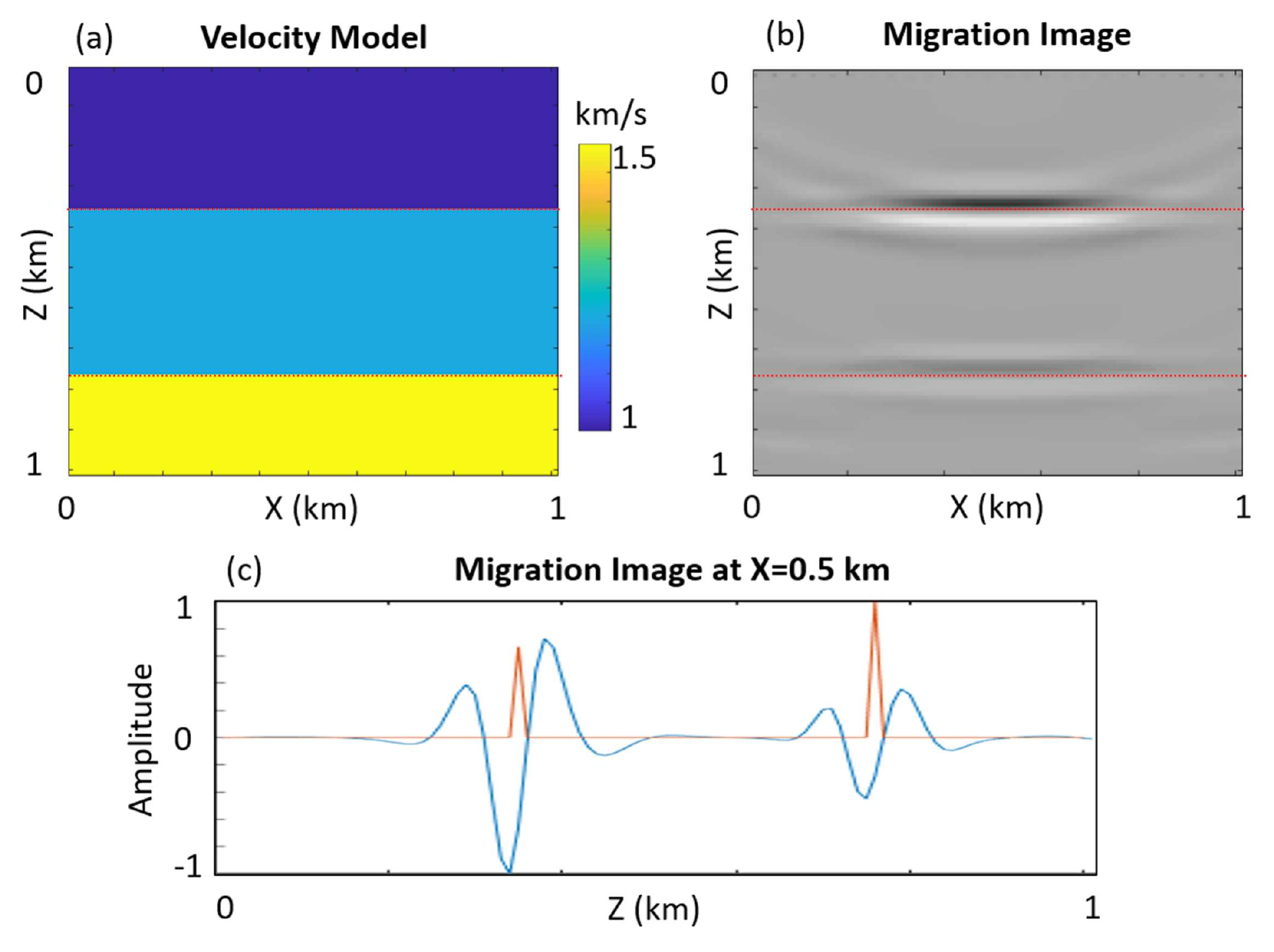}
\caption{a) Three-layer velocity model, b) RTM image, and c)
migration image (blue curve) at X=0.5 km, where the red curve
is the normalized reflectivity model.}
\label{zhaolun.fig1_1}
\end{figure}
\begin{figure}
  \centering
  \includegraphics[width=\columnwidth]{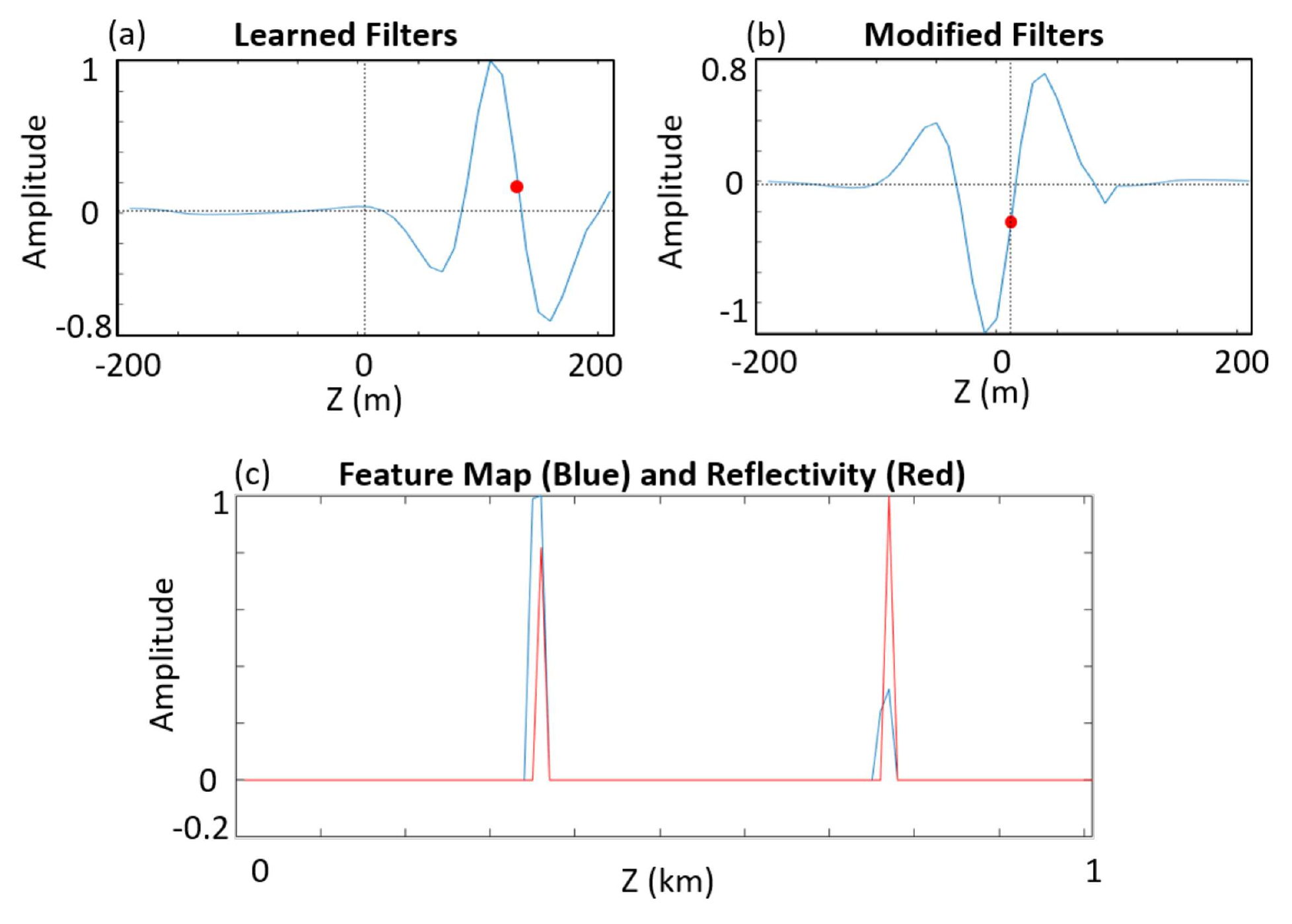}
\caption{a) Learned filter, b) modified filter,
and c) feature map (blue) and reflectivity model (red).}
\label{zhaolun.fig2_1}
\end{figure}

\subsection{Three-layer Velocity Model}

The interpretation of feature maps and
filters can be understood by computing them for the
Figure \ref{zhaolun.fig1_1}a model.
The grid size of the model is 101$\times$101, and the grid interval is
10 m in both the x and z directions. There are 26 shots evenly
spaced at a distance of 40 m on the surface, and each shot is recorded
by  101 receivers with a sampling interval of 10 m.
Figure~\ref{zhaolun.fig1_1}b show the reverse time migration (RTM)
image.

The first test is for a 1D model where
we extract the image located at X$=0.5$ km,
which is displayed as the blue curve in Figure~\ref{zhaolun.fig1_1}c.
The red curve in Figure~\ref{zhaolun.fig1_1}c is the reflectivity model.
Assume that there is only one filter in $ {\boldsymbol  \Gamma}$
and it extends over the depth of  400 m (41 grid points).
We now compute the NNLSM image by finding the optimal ${ \bf m}$
and ${\boldsymbol \Gamma}$ by the two-step iterative
procedure denoted as the alternating  descent method (see \citet{liu2018neural} and \citet{liu2018}).
The computed filter {$\bm \gamma$}$_i$ is shown in Figure~\ref{zhaolun.fig2_1}a
where the phase of the filter ${\boldsymbol  \gamma}_i $ is nonzero.
If we use a filter with a non-zero time lag to calculate its feature map ${\bf m}$,
the phases of the feature map and the true reflectivity $\m$ will be different.
So, we need to modify the time lag and polarity of the basis function
$\tilde {\boldsymbol  \Gamma}_i$. The modified basis function is
shown in Figure~\ref{zhaolun.fig2_1}b, and its coefficients are
displayed as the blue curve in Figure~\ref{zhaolun.fig2_1}c.
Compared with the true reflectivity $\m$ (red curve in
Figure~\ref{zhaolun.fig2_1}), the feature map can give the correct
positions but also give the wrong values of the reflectivity distribution.

\begin{figure}
  \centering
  \includegraphics[width=0.8\columnwidth]{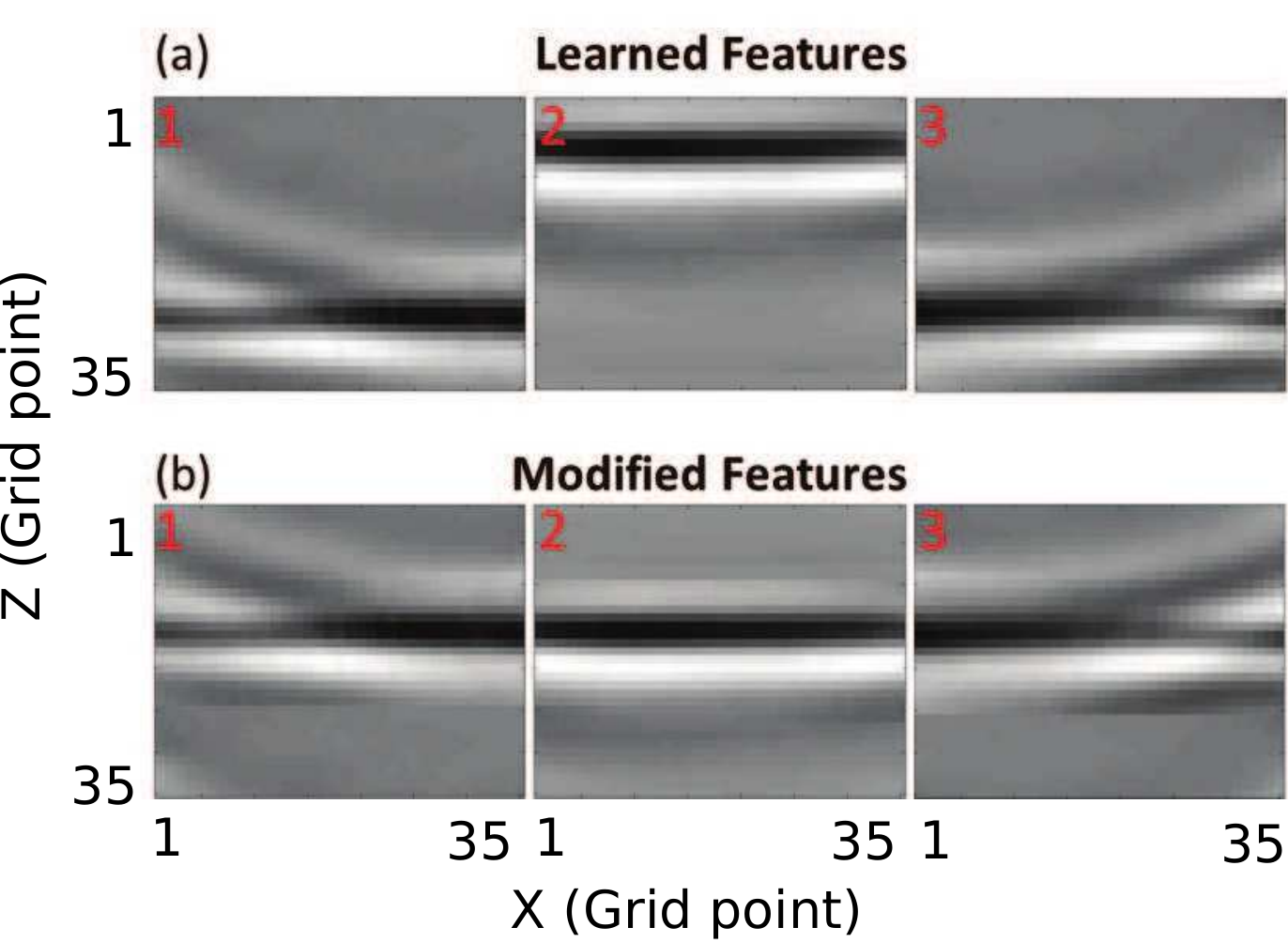}
\caption{a) Learned and b) modified features.}
\label{zhaolun.fig3}
\end{figure}

\begin{figure}
  \centering
  \includegraphics[width=0.8\columnwidth]{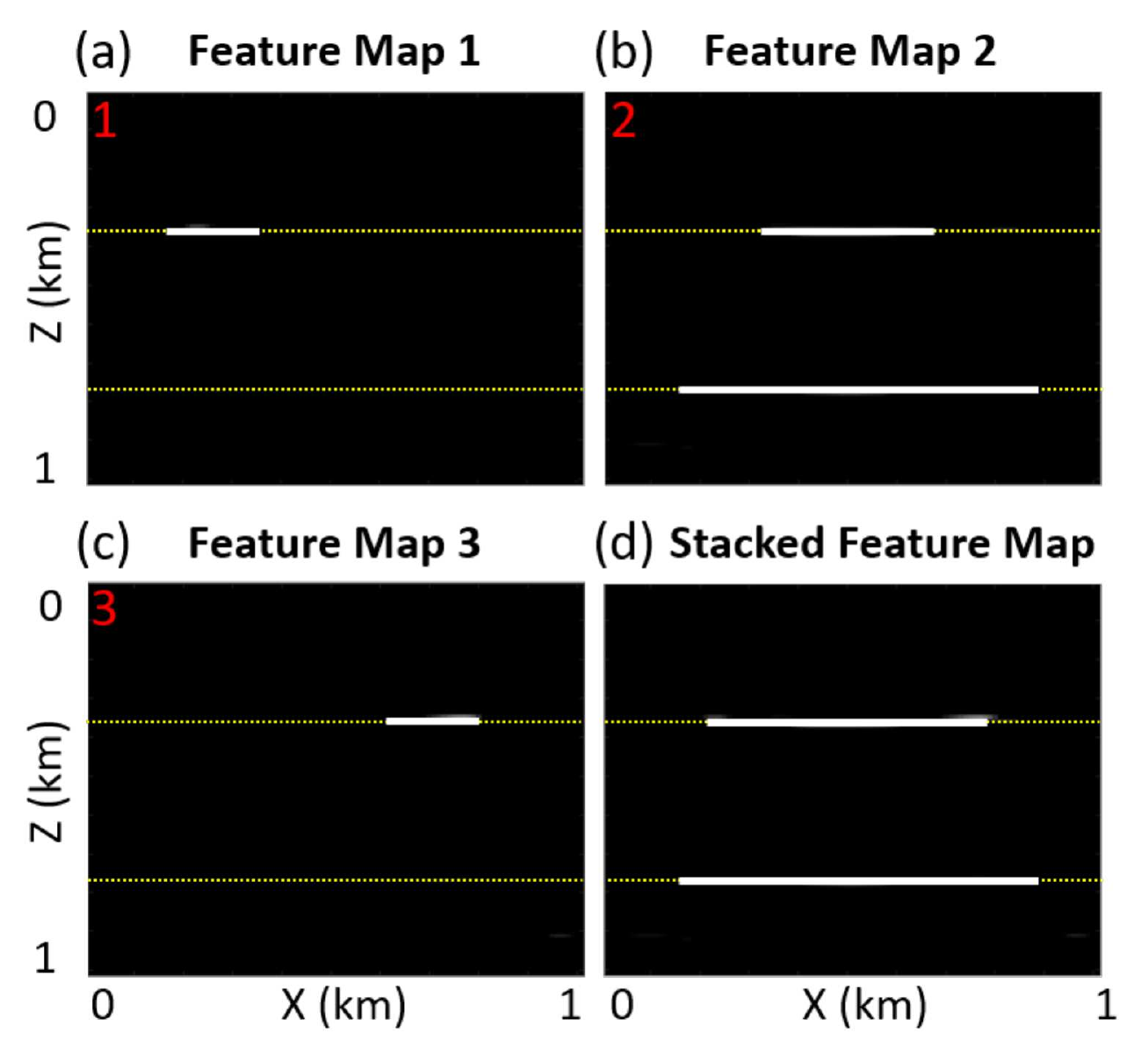}
\caption{Feature maps for the features a) 1, b) 2, and c) 3 shown
  in Figure~\ref{zhaolun.fig3}. The stacked feature map
  is shown in d). Here, the white lines show the locations of non-zero points and
the yellow lines indicate the locations of the reflectivity distributions. }
\label{zhaolun.fig4}
\end{figure}

Next, we perform a 2D test where the input is the 2D migration image in
Figure~\ref{zhaolun.fig1_1}b. Three 35-by-35 (grid point)
filters are learned (see Figure~\ref{zhaolun.fig3}a).
The modified filters are shown in Figure~\ref{zhaolun.fig3}b.
Appendix \ref{D} describes how
we align the filters by using the cross-correlation method.
The feature maps of these three filters are displayed in
Figures~\ref{zhaolun.fig4}a-\ref{zhaolun.fig4}c. Figure~\ref{zhaolun.fig4}d
shows the sum of these three feature maps.
It is evident that the stacked feature maps can estimate the correct locations
of the reflectivity spikes.


\subsection{SEG/EAGE Salt Model}

\begin{figure*}[htbp]
  \centering
  \includegraphics[width=\textwidth]{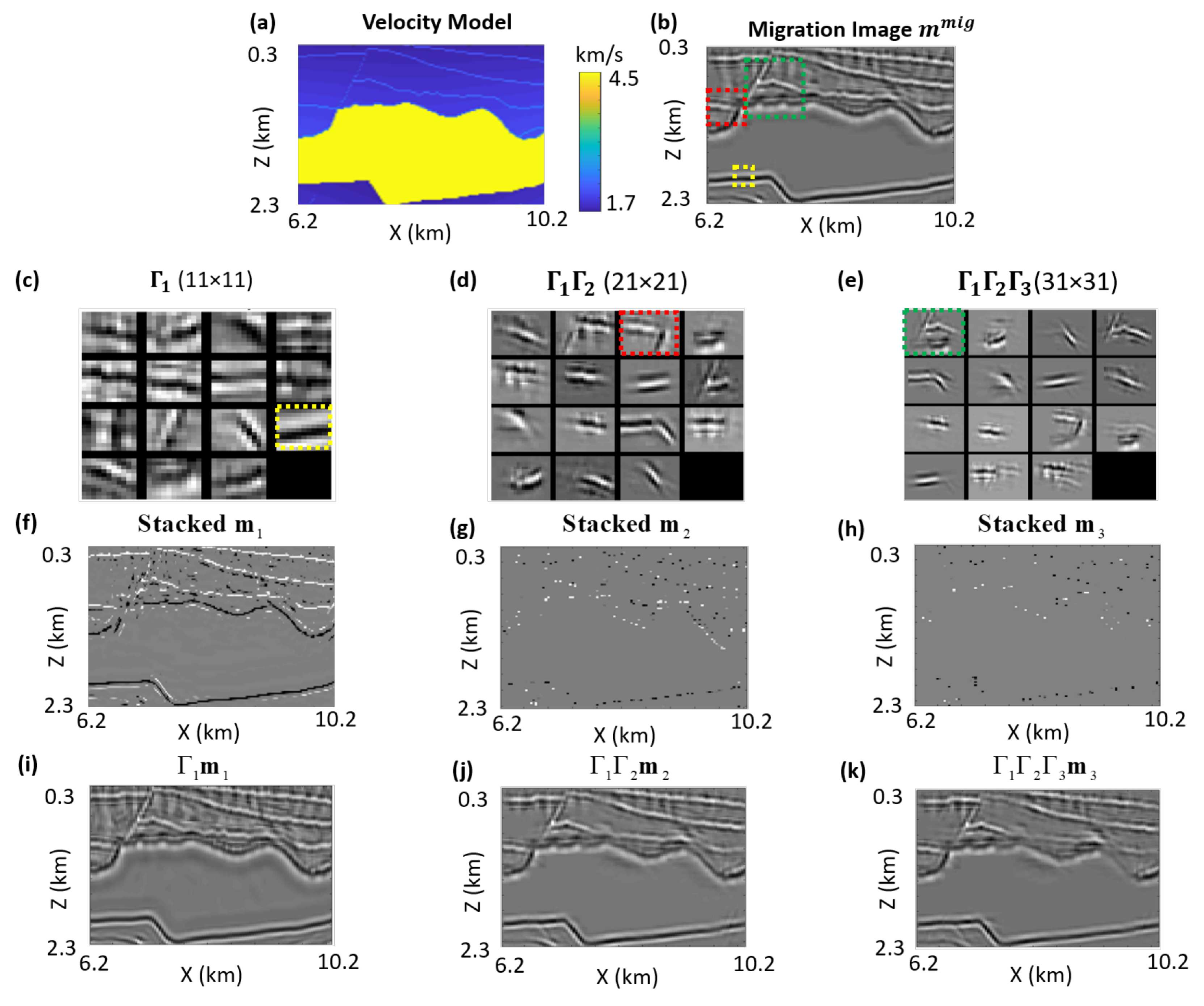}
\caption{(a) 2D SEG/EAGE salt model, (b) RTM image, (c)-(e) learned effective
  filters ${\boldsymbol \Gamma}^{(1)}$, ${\boldsymbol \Gamma}^{(2)}$
  and ${\boldsymbol \Gamma}^{(3)}$, (f)-(h) stacked quasi-reflectivity coefficients
  for $\m_1$, $\m_2$ and $\m_3$, (i)-(k) reconstructed migration
images ${\boldsymbol \Gamma}^{(1)}\m_1$, ${\boldsymbol \Gamma}^{(2)}\m_2$ and ${\boldsymbol \Gamma}^{(3)}\m_3$.}
\label{zhaolun.fig11}
\end{figure*}

\begin{figure*}[htbp]
  \centering
  \includegraphics[width=\textwidth]{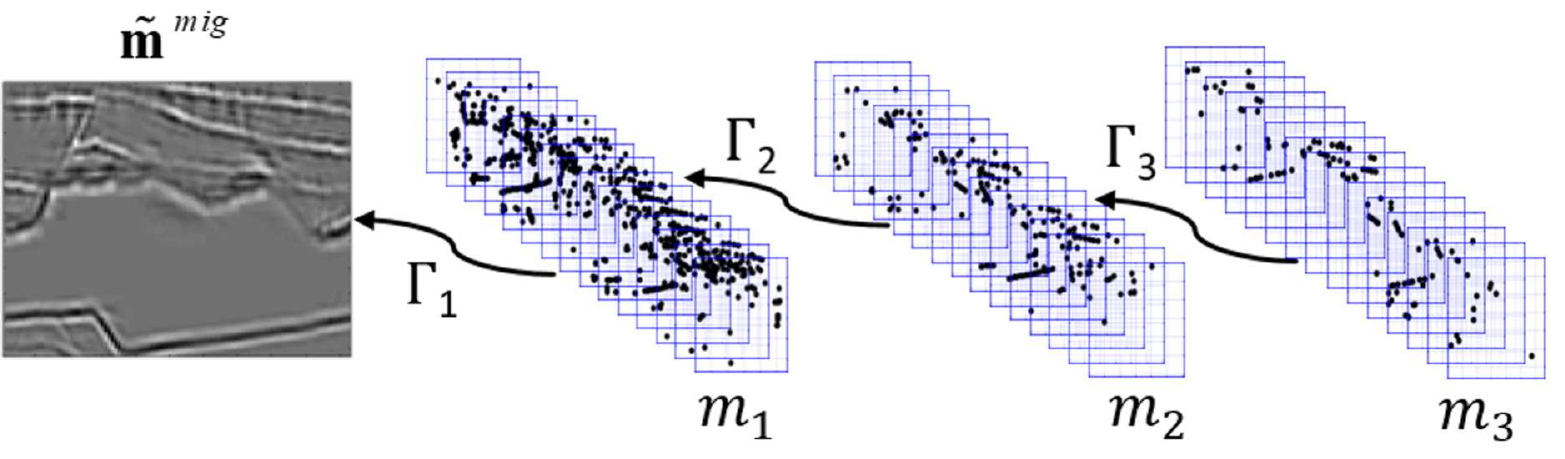}
\caption{Decomposition of the migration image of the 2D SEG/EAGE salt model in terms of the multi-layer
  filter $\boldsymbol \Gamma_i$ and pre-stacked quasi-reflectivity coefficient $\m_i$,
  where the black dots in ${\bf m}_i$ represent the nonzero value.}
\label{fig:sparsecoef}
\end{figure*}

The multilayer NNLSM procedure (see equation \ref{eqn:multilayerSLSM})
is now applied to the migration image associated with
the 2D SEG/EAGE salt velocity model in Figure \ref{zhaolun.fig11}a.
The grid size of the model is 101 grid points in both the z- and
 x-directions. The grid interval is 40 m in the x-direction
and 20 m in the z-direction.  Figure~\ref{zhaolun.fig11}b shows the
reverse time migration (RTM) image.
The multilayer NNLSM consists of three convolutional layers: the
first one contains 15 basis functions, i.e., filters,
of size 11$\times$11 grid points, the
second one consists of 15 basis functions with
dimensions 11$\times$11$\times$15, and the last one contains contains
15 basis function of dimensions 11$\times$11$\times$15.
Equation \ref{eqn:multilayerSLSM} is solved for both ${ {\bf m}_i}$ and ${\boldsymbol \Gamma}_i$ ($i \in 1,2,3$) by the two-step iterative procedure denoted as the
alternating  descent method. The multilayered structure is shown in
Figure~\ref{fig:sparsecoef}, where the black dots in ${\bf m}_i$ represent the nonzero values of the quasi-reflectivity distribution.
The effective basis functions computed for these layers are shown in
Figures~\ref{zhaolun.fig11}c-\ref{zhaolun.fig11}e, where the yellow, red and green boxes
indicate the sizes of the effective basis functions, which can
 be considered as quasi-migration Green's functions.
 It indicates that the basis
functions of the first layer ${\boldsymbol \Gamma}_1$ contain very simple
small-dimensional edges, which are called ``atoms'' by \citet{sulam2018multilayer}. The non-zeros of the second group of basis functions
${\boldsymbol \Gamma}_2$ combine a few atoms from ${\boldsymbol \Gamma}_1$ to create slightly more complex
edges, junctions and corners in the effective basis functions in ${\boldsymbol \Gamma}^{(2)}$.
Lastly, ${\boldsymbol \Gamma}_3$ combines atoms from ${\boldsymbol \Gamma}^{(2)}$ in order to reconstruct the more
complex parts of the migration image.  The corresponding stacked
coefficient images, also known as feature maps,
are shown in Figures \ref{zhaolun.fig11}f-\ref{zhaolun.fig11}h,
which give the quasi-reflectivity distributions. The reconstructed migration images are
 shown in Figures \ref{zhaolun.fig11}i-\ref{zhaolun.fig11}k.

\begin{figure}[htpb]
  \centering
  \includegraphics[width=\columnwidth]{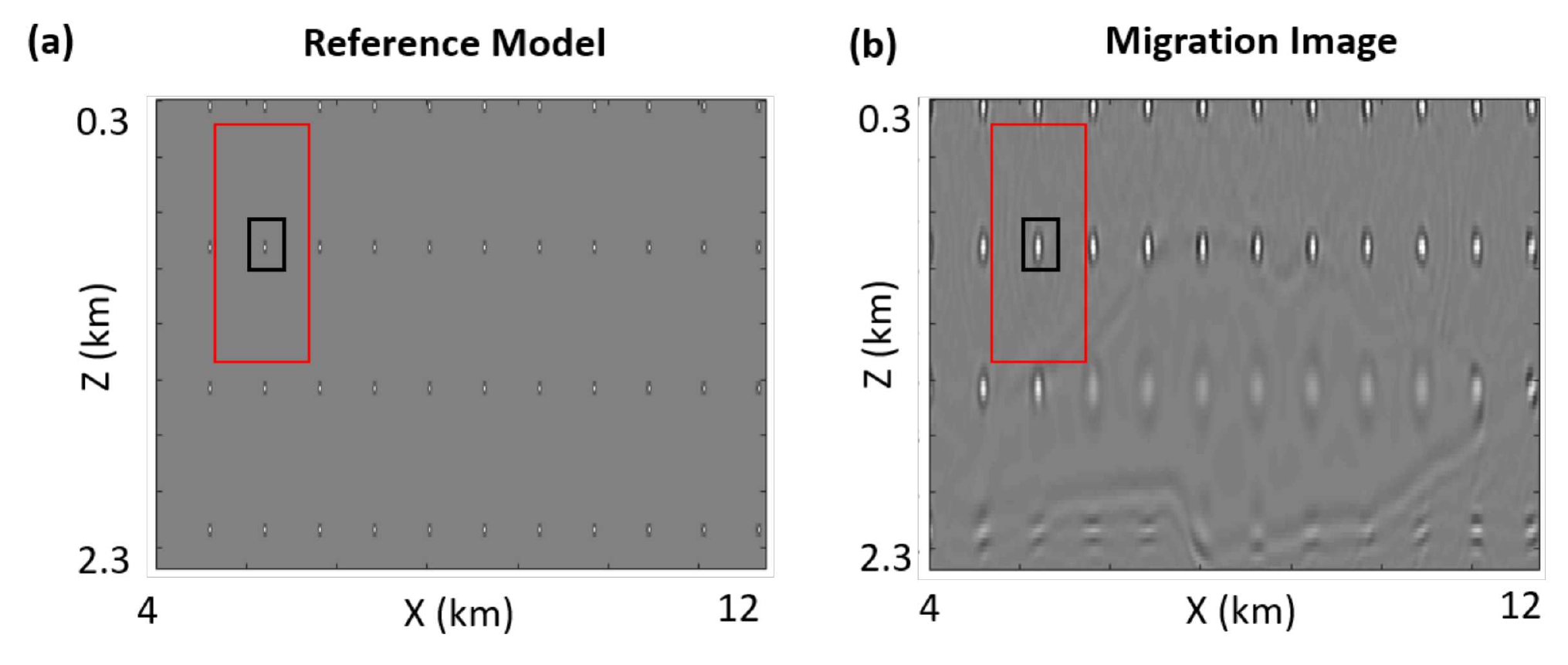}
  \caption {(a) Reference model and (b) its migration image for the standard deblurring LSM method.}
  \label{fig:mdref}
\end{figure}

\begin{figure}[htpb]
  \centering
  \includegraphics[width=\columnwidth]{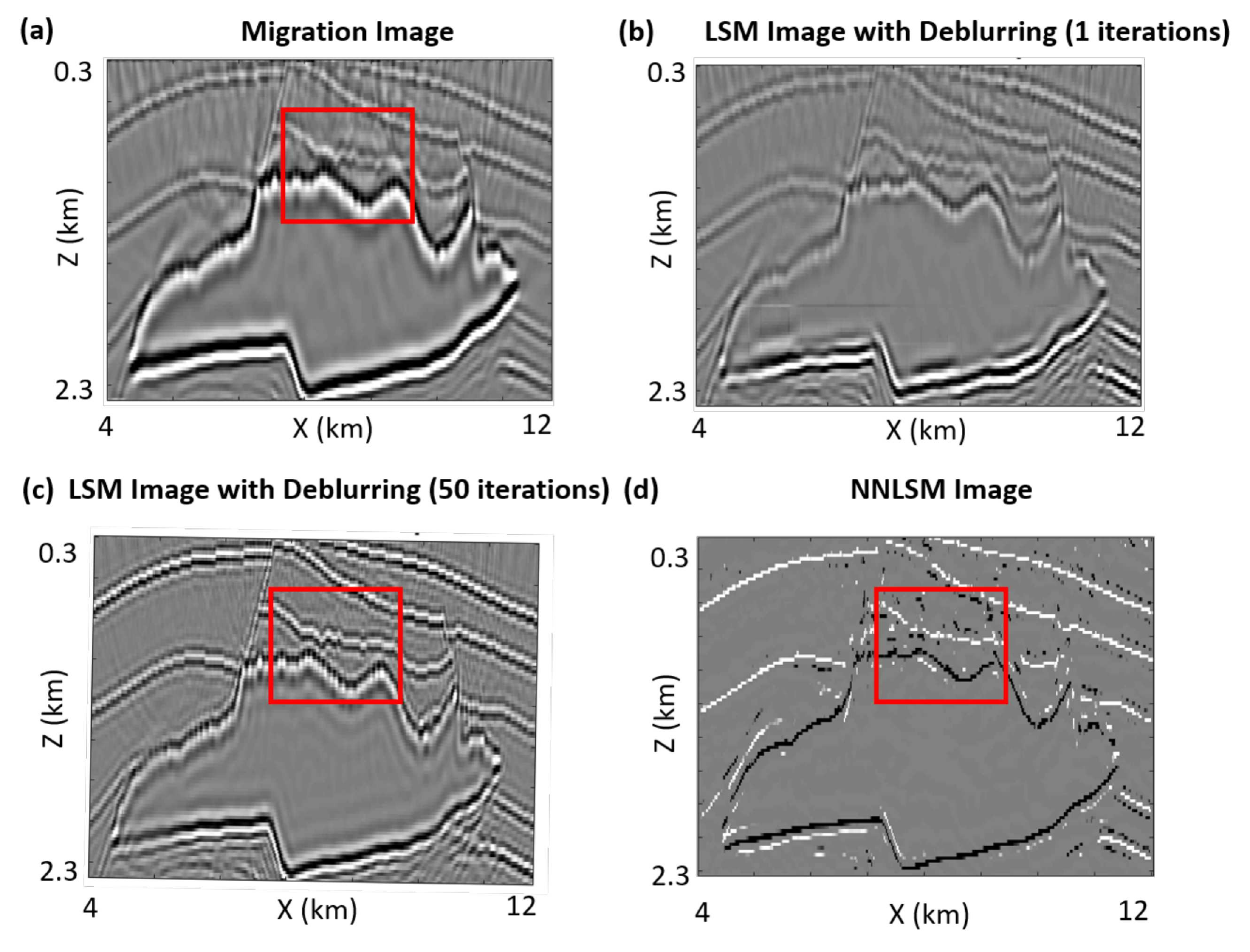}
  \caption {(a) RTM image, (b) the first and (c) $50^{th}$ iteration results by LSM with deblurring, and (d) NNLSM image.}
  \label{fig:rtmComp}
\end{figure}

For comparison, we computed the standard LSM image using the deblurring method described
in \citet{chen2017q,chen2019hybrid}. Here, the deblurring filter size is 17x17
grid points (black boxes in Figure~\ref{fig:mdref})
and computed for a 50x50 grid  (red boxes in Figure~\ref{fig:mdref})
of evenly spaced point scatterers with the same migration velocity
model as used for the data migration in Figure \ref{zhaolun.fig11}a. The
standard LSM images for the first and 50$^{th}$ iterations are shown in Figures ~\ref{fig:rtmComp}b
and ~\ref{fig:rtmComp}c, respectively, next to the NNLSM image in Figure~\ref{fig:rtmComp}d.
It is clear that the NNLSM image is better resolved than the LSM image,
although there are discontinuities in some of the NNLSM interfaces not seen in the LSM image.
Some of the detailed geology is lost in the LSM image as seen in the wiggly interface
in the red-rectangular area of  Figure~\ref{fig:rtmComp}.  The practical application of
the NNLSM image
is that it might serve a super-resolved attribute image that can be combined with other attributes
to delineate geology. For example, combining the depth-slice of the NNLSM image with
a spectral decomposition image \citep{spectral} can help delineate the lithological
edges of meandering channels.

\begin{figure}[htpb]
  \centering
  \includegraphics[width=\columnwidth]{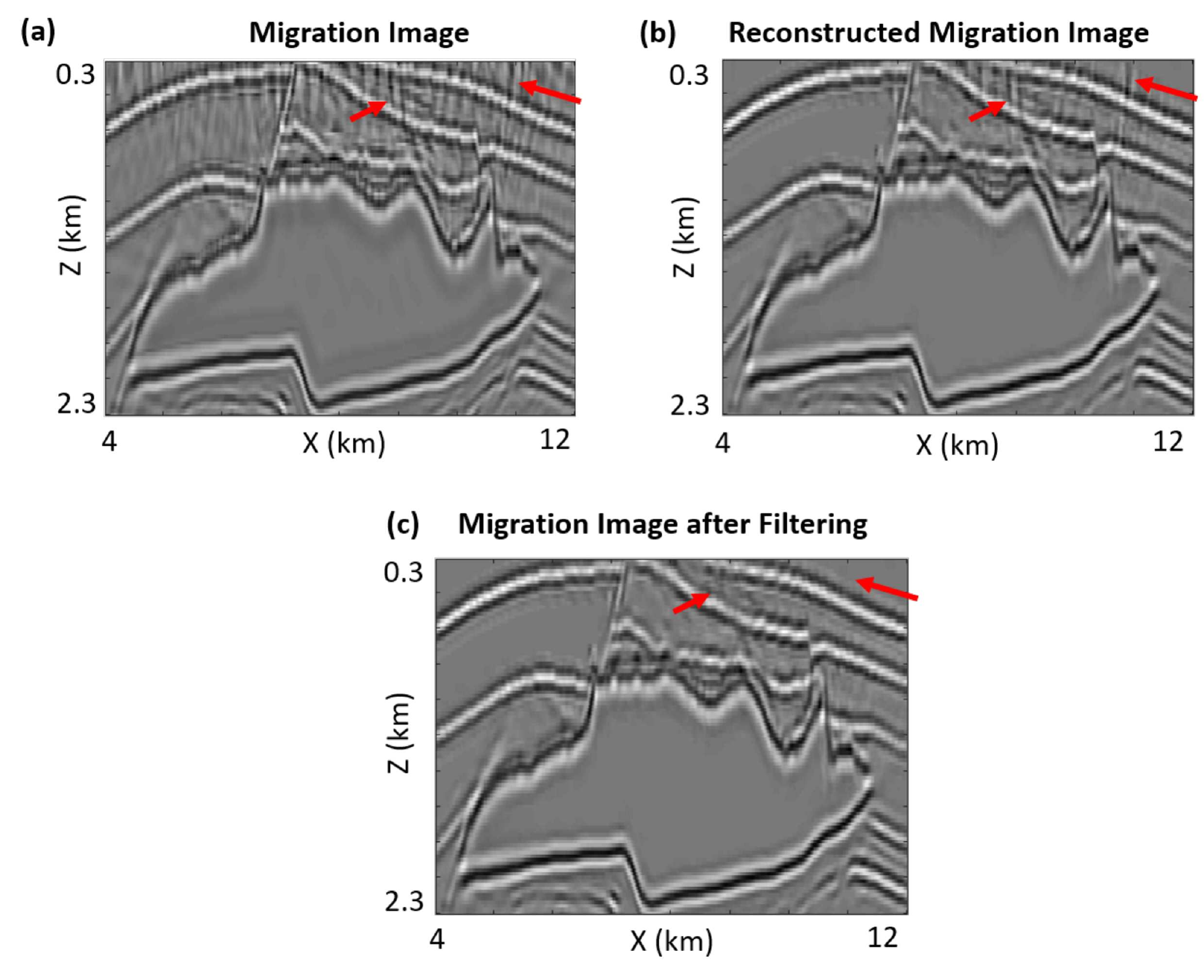}
  \caption {(a) RTM image, (b) reconstructed image with all basis functions, and (c) reconstructed image with selected
  basis functions.}
  \label{fig:filtering}
\end{figure}

\begin{figure}[htpb]
  \centering
  \includegraphics[width=\columnwidth]{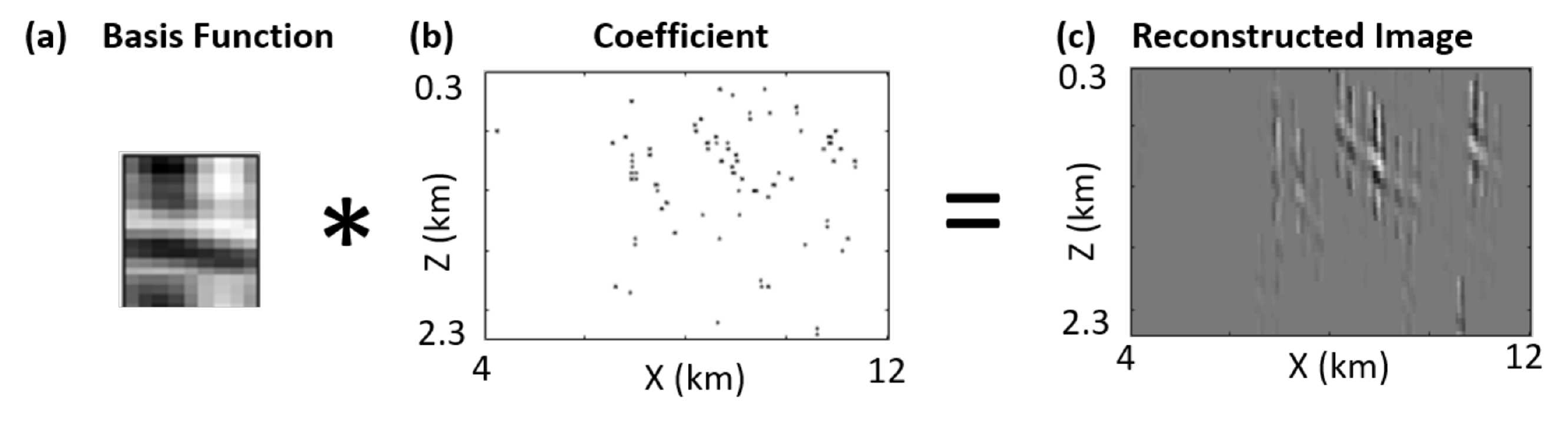}
  \caption {(a) basis function from noise, (b) the coefficients, and (c) reconstructed image.}
  \label{fig:noisefilter}
\end{figure}

{NNLSM can filter out random and coherent noises in the migration image
after reconstructing the migration image by eliminating the noisy learned basis functions
and their coefficients in the NNLSM image.  For example, Figure~\ref{fig:filtering}a
shows the RTM image with a sparse acquisition geometry so that the image contains a
strong acquisition footprint.  The reconstructed migration image in Figure~\ref{fig:filtering}b
shows significant mitigation of this noise in Figure \ref{fig:filtering}a.
However, the migration swing noise is still prominent near 
the red arrows in Figure \ref{fig:filtering}b.
Such noise is reconstructed from the noisy basis function shown
in Figure~\ref{fig:noisefilter}a and the coefficients in Figure~\ref{fig:noisefilter}b.
Figure~\ref{fig:noisefilter}c is the image reconstructed by correlating the basis
function in Figure \ref{fig:noisefilter}a with the coefficients in Figure~\ref{fig:noisefilter}b.
After filtering out the basis functions from noise, the reconstructed image is
shown in Figure~\ref{fig:filtering}c, which is free from aliasing noise at the locations indicated by the red
 arrows.}

\subsection{North Sea Data}

\begin{figure*}[htbp]
  \centering
\includegraphics[width=\textwidth]{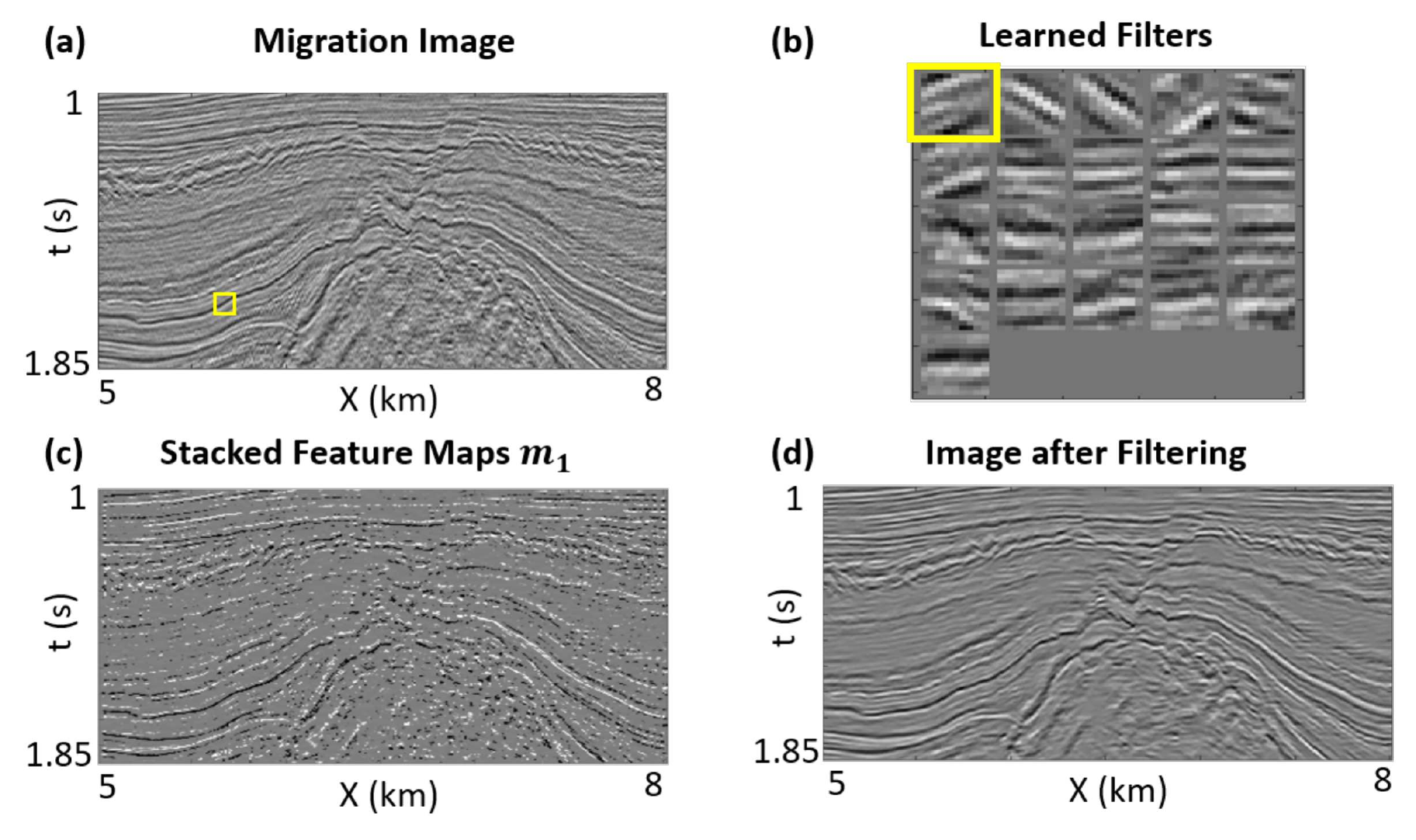}
\caption{a) Migration image computed from the F3 offshore block data, b) learned filters,
  c) stacked feature maps and d) migration image after filtering.}
\label{zhaolun.fig2}
\end{figure*}

We apply the NNLSM method to  field data collected in  the
 North Sea~\citep{schroot_schuttenhelm_2003}, where the time migration
 image is shown in Figure~\ref{zhaolun.fig2}a.
The time axis is gridded with 213 evenly-spaced points and there
are 301 grid points along the x-axis.  We compute 21 13-by-5 (grid point)
convolutional basis functions, i.e.
filters $\boldsymbol \gamma_i$ for ($i=1,2,...21$), by the
 NNLSM procedure (see Figure~\ref{zhaolun.fig2}b).
 These filters approximate the dip-filtered  migration Green's functions, and the
basis function is marked as the yellow boxes in Figure~\ref{zhaolun.fig2}a and \ref{zhaolun.fig2}b.
The stacked feature maps (quasi-reflectivity distribution) are displayed in Figure~\ref{zhaolun.fig2}c.
It is evident that the stacked feature maps can provide a high-resolution migration image.
After reconstruction from the learned filters and feature maps, the migration image
is shown in Figure~\ref{zhaolun.fig2}d with less noise.

\begin{figure*}[htbp]
  \centering
\includegraphics[width=\textwidth]{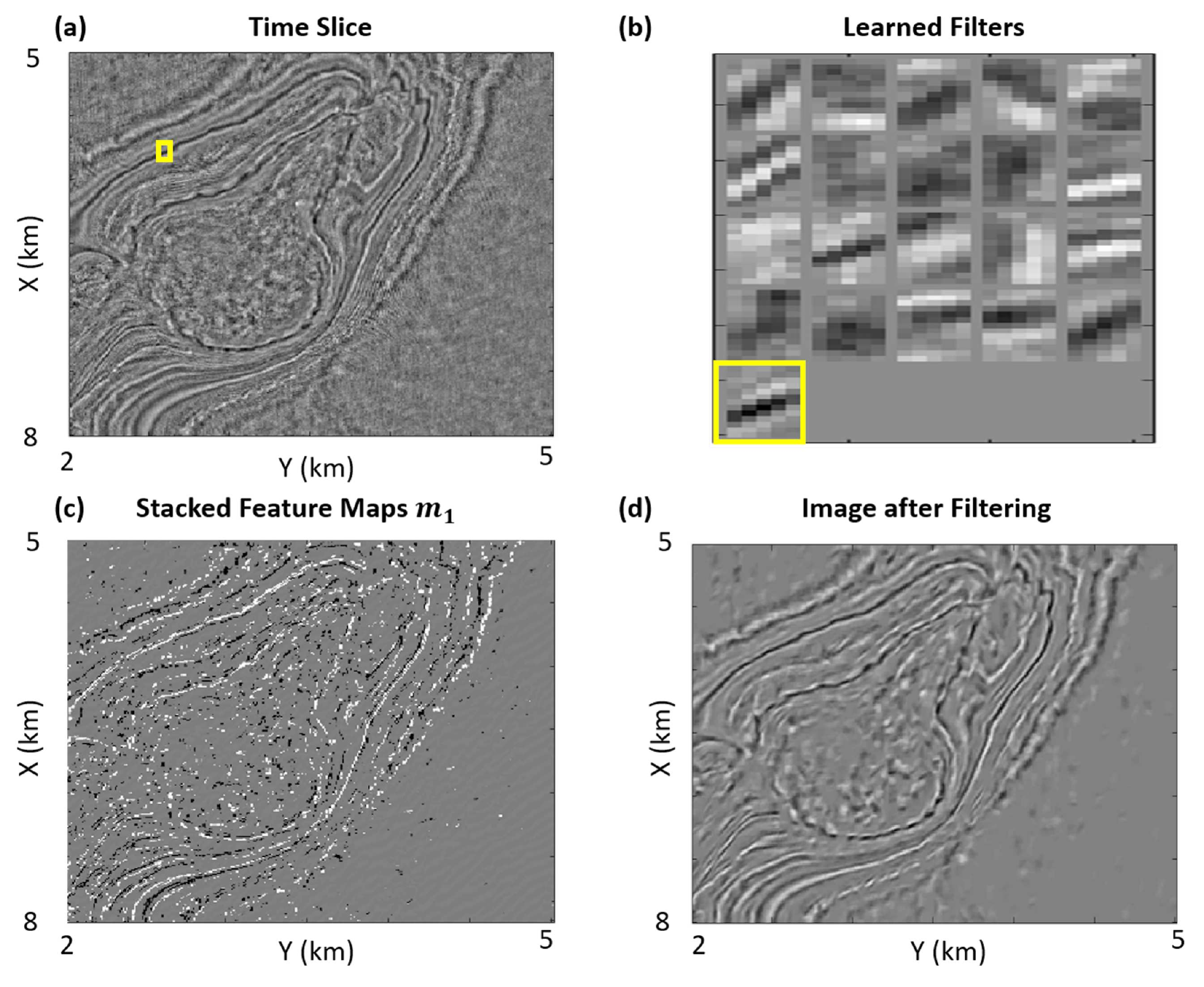}
\caption{a) Time slice of the migration image computed from the F3 offshore block data,
  b) learned filters,
  c) stacked feature maps and d) migration image after filtering.}
\label{fig:fieldTimeslice}
\end{figure*}

Finally, we apply NNLSM to a time slice of the migration image,
which is shown in Figure~\ref{fig:fieldTimeslice}a, and the image size is 301 by 301 gridpoints.
Figure~\ref{fig:fieldTimeslice}b shows the 21 13$\times$5 filters estimated
by the NNLSM procedure. The stacked feature map is displayed in Figure~\ref{fig:fieldTimeslice}c,
which may be used as a superresolution attribute image for high-resolution delineation of geologic bodies.
The reconstructed migration image is shown in Figure~\ref{fig:fieldTimeslice}d and we can see there is
less noise.

\begin{figure}[htpb]
  \centering
  \includegraphics[width=\columnwidth]{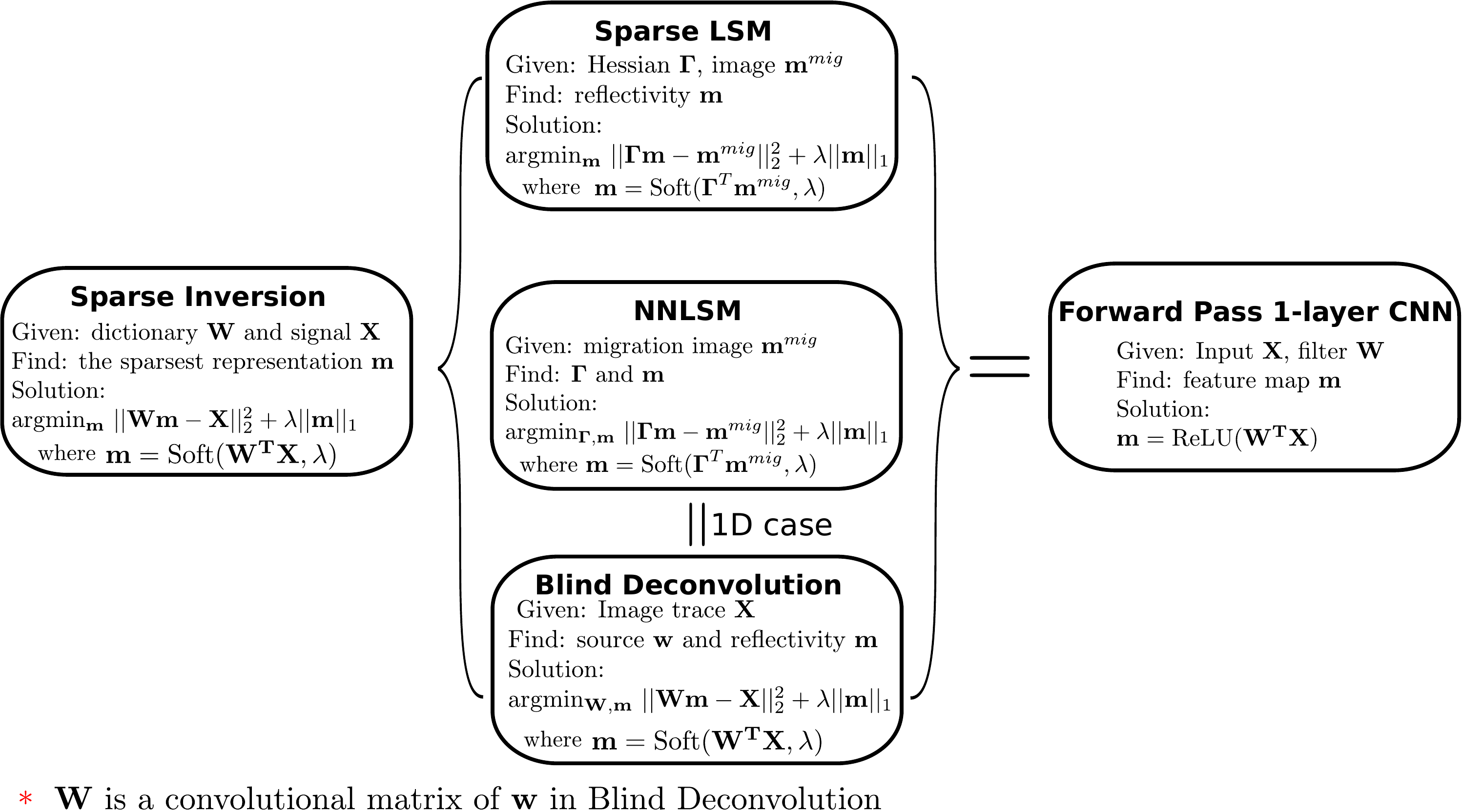}
  \caption {The relationship between sparse inversion and CNN, where the sparse inversion methods include
  sparse LSM, NNLSM, and blind deconvolution.}
  \label{fig:lsmcnn}
\end{figure}

\section{Discussion}

The forward modeling for a multilayered neural network
 is shown to be equivalent to a single-iterate solution of a multilayered LSM problem.
 This assumes positivity and shrinkage constraints on the soft thresholding operation,
 so it reduces to the ReLu operation.  This equivalence relates the physics of seismic imaging to architectural
features in the neural network.
\begin{itemize}
\item The size of the filters in the first layer should be about the same size as the Green's function for that model. Experiments with numerical models suggest that this size is approximately one-to-two wavelengths. In this case, the filter is interpreted as an approximation to the migration Green's function, except it
is that for a reflecting segment. Thus, we interpret the approximate migration Green's function as a migration segment spread function (SSF) rather than
a migration point spread function.  \citet{sulam2018multilayer} classifies each feature in the first layer  as an atom which takes on the role of a SSF.
\item The output of the first layer provides the small scale,
i.e., high-wavenumber, features associated with the input data.
For an input migration image, the
feature maps of the first layer resemble sub-quasi-reflectivity maps
of the subsurface.  Adding the
sub-quasi-reflectivity maps together gives a close approximation to the actual reflectivity model as shown in Figures~\ref{zhaolun.fig4}d and \ref{zhaolun.fig11}f.

\item
The output of the second layer is a weighted sum of the first-layer features,
which create sparser feature maps. \citet{sulam2018multilayer} classifies the concatenation of the
filters from the first and second layers  as molecules (see equation
\ref{eqn:effectiveFilter}). In the migration problem,
 the resulting filters are SSFs for even larger segments of the original
 reflector boundaries. The feature maps of
the third layer are a weighted sum of the second layer's
features to produce even the sparsest feature maps.
 For migration, the final feature maps are very sparse while the
concatenated filters are associated with large-scale features of the migration image.

\item The computational cost of computing NNLSM images is significantly 
  less than that for LSM images
  because no solutions of the wave equation are needed.  For example, we consider  
  the 2D FDTD forward modeling of the acoustic equation with
  an eighth-order scheme in space and a second-order scheme in time, and its computational complexity 
  is $O(n_tn^2)$ for one shot, where $n$ is the number of grid points
  in one direction and $n_t$ is the number of the time steps. According to \citet{Plessix2007}, 
  $n_t$ is approximate $30\ n$ to satisfy the stability condition and to make sure there is enough 
  recoding time when $v_{max}/v_{min}\simeq=3$, where $v_{max}$ and $v_{min}$ are 
  the maximum and minimum velocities, respectively. So,  The complexity of 2D FDTD forward modeling
  of the acoustic equation is $O(n_sn^3)$ where $n_s$ is the number of the shots.
  The complexity of LSRTM is $O(6n_sN^{iter}n^3)$ \citep{schuster2017seismic},
  where $N^{iter}$ is the iteration number. 
  For NNLSM, the complexity is $O(N^{iter}n^2\log n)$ according to \citet{heide2015fast} and
  can be reduced to $O(N^{iter}n^2)$  if a local block coordinate-descent algorithm
  is used \citep{zisselman2019local}.
\end{itemize}

The 1D NNLSM can be interpreted as a blind deconvolution (BD) problem in seismic data processing
\citep{kaaresen1998multichannel, Pawan2018}.  It can be seen in Figure~\ref{zhaolun.fig3} that
the filter of NNLSM is the source wavelet of BD
and the coefficients of NNLSM are the quasi-reflectivity coefficients of BD.
However, NNLSM can have more than one filter and the filters can be 2D or 3D 
filters.  We show the relationship between BD,  sparse LSM, NNLSM, and CNN  
in Figure~\ref{fig:lsmcnn}.

NNLSM is an unsupervised learning method. Compared to a supervised
learning method, it does not heavily depend on the huge amount
of training data. However, it may need human intervention for inspecting the
quality of the quasi-reflectivity images and we need to align the filters to get
more consistent quasi-reflectivity.

In Figure~\ref{fig:fieldTimeslice}, we apply 2D NNLSM to a time slice of the 3D North Sea Data.
The 2D NNLSM method can be extended to a 3D implementation to learn a set of 3D filters. 
Incorporating the third dimension of information from 3D filters will likely lead to a better 
continuity of the reflectors in Figure~\ref{fig:fieldTimeslice}c.
However, the computational cost will increase by more than a multiplicative factor of $n$.

\section{Conclusion}
Neural network least squares migration finds
the optimal quasi-reflectivity distribution and
quasi-migration-Green's functions  that minimize
 a sum of migration misfit and  sparsity regularization functions.
 The advantages of NNLSM over standard LSM are that
its computational cost is significantly less than that for LSM and
it can be used for filtering
both coherent and incoherent noise in  migration images.
A practical application of the NNLSM image is as an attribute map that provides superresolution imaging
of the layer interfaces. This attribute image can be combined with other attributes to
delineate both structure and lithology in depth/time slices of migration images.
Its disadvantage is that the NNLSM quasi-reflectivity image is
only an approximation to the actual reflectivity distribution.

A significant contribution of our work is
that we show that the filters and feature
maps of a multilayered CNN are analogous to the
migration Green's functions and reflectivity distributions. For the first time we now have
a physical interpretation of the filters and feature maps in deep CNN in terms
of the operators for seismic imaging.  Such an understanding has the potential to lead
to better architecture design of the network and extend its application to waveform inversion.
In answer to Donoho's plea for more rigor, NNLSM represents a step forward in establishing
the mathematical foundation of CNN in the context of least squares migration.

\section{Acknowledgements}
The research reported in this publication was supported by the King Abdullah University of Science and Technology (KAUST) in Thuwal, Saudi Arabia. We are grateful to the sponsors of the Center for Subsurface Imaging and Modeling Consortium for their financial support. For computer time, this research used the resources of the Supercomputing Laboratory at KAUST and the IT Research Computing Group. We thank them for providing the computational resources required for carrying out this work.

\append{Migration Green's Function}

\citet{schuster2000green} show that the poststack migration \citep{yilmaz2001seismic}
 image $m(\x)^{mig}$ in the frequency domain is computed by weighting each reflectivity
 value $m(\z)$ by $\Gamma(\x|\z)$ and integrating over the model-space coordinates $\z$:
 \begin{eqnarray}
 &{\mbox for} ~\x \in D_{model}:~~~~m(\x)= \nonumber \\
 &\eta \int_{D_{model}} d\z \overbrace{\int_{\y \in D_{data}} d\y \omega^4 G(\x|\y)^{2*}G(\y|\z)^2 }^{\Gamma(\x|\z)}m(\z),
 \label{SLSM.AAeq5aa}
 \end{eqnarray}
 where $\eta$ represents terms such as the frequency variable raised to the 4th power.
The migration Green's function $\Gamma(\x|\z)$ is
given by
\begin{eqnarray}
&{\mbox for} ~\x,\z \in D_{model}: ~~~~\Gamma(\x|\z) =\nonumber \\
&\eta \int_{\y \in D_{data}} d\y \omega^4 G(\x|\y)^{2*}G(\y|\z)^2 .~~~~~~
\label{SLSM.AAeq4}
\end{eqnarray}
Here we implicitly assume a normalized source wavelet
in the frequency domain, and
 $D_{model}$
and $D_{data}$ represent the sets of coordinates in, respectively,
the model and data spaces. The term $G(\x'|\x)=e^{i\omega \tau_{xx'}}/||\x-\x'||$ is the
Green's function for a source at $\x$ and a receiver at $\x'$
in a smoothly varying medium\footnote{If the source and receiver are coincident at $\x$ then the zero-offset trace
is represented by the squared Green's function $G(\x|\x')^2$.}. The traveltime
 $\tau_{xx'}$ is for a direct arrival to propagate from $\x$ to $\x'$.

The physical interpretation of the kernel $\Gamma(\x'|\x)$
is that it is the migration operator's\footnote{This assumes that the zero-offset trace
is generated with an impulsive point source
with a smoothly varying background velocity model,
and then migrated by a poststack migration operation.
It is always assumed that the direct arrival
is muted and there are no multiples.}
 response at $\x'$ to
a point scatterer at $\x$, otherwise known as the MGF or the migration
Green's function \citep{schuster2000green}.
It is analogous to the point spread function (PSF)
of an optical lens for a point light source at $\x$ in front
of the lens and its optical image  at $\x'$ behind the lens on the image plane.
In discrete form, the modeling term
$[{\boldsymbol \Gamma \m}]_i$ in equation~\ref{SLSM.AAeq5aa} can be expressed as
\begin{eqnarray}
[{\boldsymbol \Gamma}\m]_i&=&\sum_j \Gamma(\x_i|\z_j) m_j.
\label{SLSM.AAeq5}
\end{eqnarray}
with the physical interpretation that $[{\boldsymbol \Gamma}\m]_i$
is the migration Green's
 function response at $\x_i$. An alternative interpretation
 is that $[{\boldsymbol \Gamma}\m]_i$ is the weighted sum of basis functions
$\Gamma(\x_i|\z_j)$ where the weights are the
reflection coefficients $m_j$ and the summation is over the $j$ index.
We will now consider this last interpretation and redefine the problem as finding
both the weights $m_i$ and the basis functions $\Gamma(\x_i|\z_j)$. This will be
shown to be equivalent to the
problem of a fully connected (FCN) neural network.

\append{Soft Thresholding Function}
Define the sparse inversion problem  as finding
the optimal value $x^*$ that minimizes the objective
function
\begin{eqnarray}
\epsilon &=&\frac{1}{2} ||\z-\x||_2^2 + \lambda ||\x||_1,
\label{SLSM.Aeq1}
\end{eqnarray}
where the $L_1$ norm demands sparsity
in the solution $\x$.
An example is where $\z$ is a noisy $M\times N$  image
such that $\z=\x+\mbox{noise}$, and we seek the optimal vector
$\x$ that satisfies equation~\ref{SLSM.Aeq1}.
Here, the noisy $M\times N$ image has been flattened
into the tall $MN\times 1$ vector $\z$.

The stationary condition for equation~\ref{SLSM.Aeq1}
is
\begin{eqnarray}
\frac{\partial \epsilon}{\partial x_i} &=& (x_i-z_i) + \lambda \frac{\partial ||\x||_1}{\partial x_i},\nonumber\\
&=&0,
\label{SLSM.Aeq2}
\end{eqnarray}
where
\begin{eqnarray}
\frac{\partial ||\x||_1}{\partial x_i}=1~for~x_i \ge 0;&
\frac{\partial ||\x||_1}{\partial x_i }=-1~for~x_i<0.
\label{SLSM.Aeq3}
\end{eqnarray}
Equations~\ref{SLSM.Aeq2}-\ref{SLSM.Aeq3} can be combined to give
the optimal $x^*$ expressed as the
two-sided ReLu function
\begin{equation}
x_i=\mbox{soft}(z_i,\lambda)=
{\left\{
\begin{array}{cl}
z_i-\lambda&if~~z_i~~\ge~~\lambda\\
0& if~~|z_i| < ~~\lambda \\
z_i+\lambda&if~~z_i~~<  -\lambda
\end{array}
\right\}
}.
\label{SLSM.Aeq4}
\end{equation}

More generally,  the iterative-soft-threshold-algorithm (ISTA)
that finds $\x^*$
\begin{eqnarray}
\x^* &=&\arg \min_{\x} {\Big [} \frac{1}{2} ||\z-{\bf W} \x||_2^2 + \lambda ||\x||_1
{\Big ]},
\label{SLSM.Aeq10}
\end{eqnarray}
is
\begin{eqnarray}
x_i^{(k+1)} &=& {\mbox {soft}}{\Big (}\x^{(k)} - \frac{1}{\alpha} {{\bf W}}^T({\bf W} \x^{(k)}-\z),\frac{\lambda}{\alpha} {\Big )}_i.
\label{SLSM.Aeq10a}
\end{eqnarray}
There are several more recently developed algorithms that have
faster convergence properties than ISTA. For example, FISTA (Fast-ISTA) has quadratic
convergence \citep{beck2009fast}.
\append{Neural Network Least Squares Migration}

The neural network least squares migration (NNLSM)
algorithm in the image domain is defined as solving for {\it both} the basis functions $\tilde \Gamma(\x_i|\x_j)$ and $\tilde m_j$ that minimize the objective function defined in
equation~\ref{SLSM.eq1}.
In contrast, SLSM only finds the least squares migration image in the image domain
and uses the pre-computed migration Green's functions that solve the wave equation.

The NNLSM solution is defined as
\begin{eqnarray}
(\tilde \m^*,\tilde {\boldsymbol  \Gamma}^*)=\arg\min_{\tilde \m,~\tilde {\boldsymbol  \Gamma}} {\Big [} \frac{1}{2} ||\tilde {\boldsymbol  \Gamma}\tilde \m - \m^{mig} ||_2^2 + \lambda S(\tilde \m) {\Big ]},
\label{SLSM.eq7}
\end{eqnarray}
where now both $\tilde {\boldsymbol \Gamma}^*$ and $\tilde \m^*$ are
to be found. The functions with tildes are mathematical
 constructs that are not necessarily identical
 to those based on
 the physics of wave
 propagation as in  equation~\ref{SLSM.eq1}.

The explicit matrix-vector form of the objective function in equation~\ref{SLSM.eq7} is given by
    \begin{eqnarray}
\epsilon=\frac{1}{2} \sum_i {\Big [}\sum_j \tilde \Gamma(\x_i|\z_j)\tilde m_j -m_i^{mig} {\Big ]}^2 + \lambda S(\tilde \m).\label{SLSM.eq8}
\end{eqnarray}
and its Fr\'echet derivative with respect to $\tilde \Gamma(\x_{i'}|\z_{j'})$ is
given by
    \begin{eqnarray}
 {\frac{\partial \epsilon}{\partial { \tilde \Gamma}(\x_{i'}|\z_{j'})}}&=& \sum_j
   (\tilde \Gamma (\x_{i'}|\z_j)\tilde m_j-\tilde m_{i'}^{mig})\tilde m_{j'}.
   \label{SLSM.eq8a}
    \end{eqnarray}
The iterative solution of equation~\ref{SLSM.eq7}
is given in two steps \citep{olshausen1996emergence}.
\begin{enumerate}
\item Iteratively estimate $\tilde m_i$
by the gradient descent formula used with SLSM:
    \begin{eqnarray}
    \tilde m_i^{(k+1)} =  \tilde m_i^{(k)} - \alpha [\tilde {\boldsymbol \Gamma}^T (\tilde {\boldsymbol \Gamma}
    \tilde \m - \m^{mig})]_i - \lambda S(\tilde \m)_i'.
    \label{SLSM.eq2aa}
    \end{eqnarray}
    However, one migration image $\m^{mig}$
    is insufficient to find so many unknowns.  In this case
    the original migration image is broken up into many small pieces so that there are many migration images
    to form examples from a large training set.
    For prestack migration, there will be many examples
	of prestack migration images, one for each shot, and the compressive sensing technique denoted as VISTA \citep{ahmad2015variable} is used for the calculations.
\item Update the basis functions $\tilde \Gamma(\x_i|\z_j)$
    by inserting equation~\ref{SLSM.eq8a} into the gradient descent formula to get
    \begin{eqnarray}
    &\tilde \Gamma(\x_{i'}|\z_{j'})^{(k+1)}=\tilde \Gamma(\x_{i'}|\z_{j'})^{(k+1)}
    -\alpha \frac{\partial \epsilon}{\partial {\tilde \Gamma}(\x_{i'}|\z_{j'})},~ ~~~\nonumber\\
    \nonumber\\
 &=\tilde \Gamma(\x_{i'}|\z_{j'})^{(k+1)}
    \nonumber \\
&	-\alpha  ( {\Big \{ }\sum_j \tilde \Gamma (\x_{i'}|\z_j)\tilde m_j {\Big \} } - m_{i'}^{mig})\tilde m_{j'}.
    \end{eqnarray}
    It is tempting to think of $\tilde \Gamma(\x|\x')$ as the migration Green's function and $\tilde m_i$ as the component of reflectivity. However, there is yet no justification to submit to this temptation and so we must consider, unlike in the SLSM algorithm, that $\tilde \Gamma(\x|\x')$  is a sparse
    basis function and $\tilde m_i$ is its coefficient. To get the
    true reflectivity then we should equate equation~\ref{SLSM.AAeq5} to $\sum_j \tilde \Gamma(\x_i,\x_j) \tilde m_j$ and solve for $m_j$.
\end{enumerate}

\append{Alignment of the Filters}

\begin{figure}[htpb]
  \centering
  \includegraphics[width=\columnwidth]{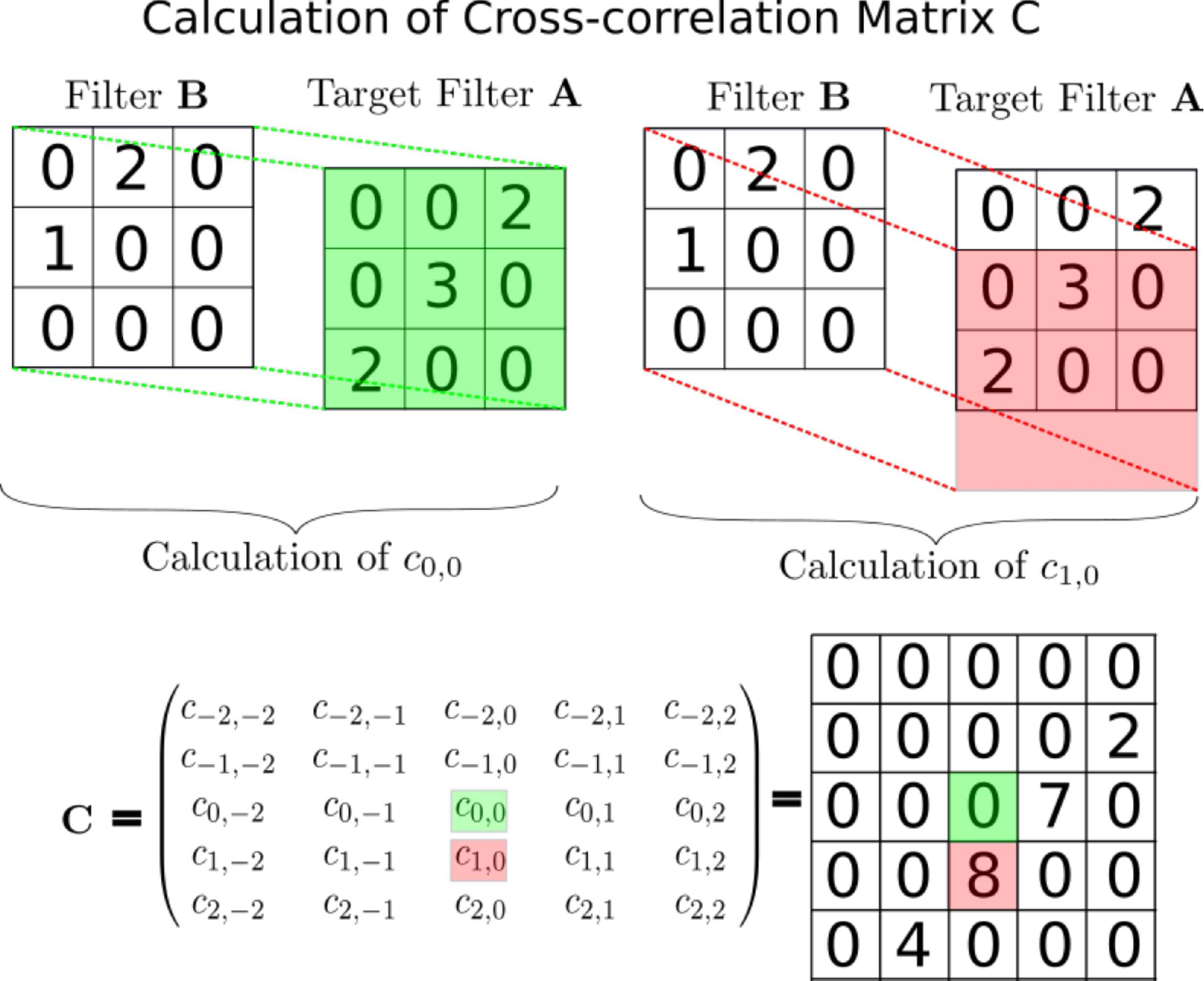}
  \caption {Calculation of the cross-correlation matrix $\bf C$.}
  \label{fig:crossCorrelation}
\end{figure}

\begin{figure}[htpb]
  \centering
  \includegraphics[width=\columnwidth]{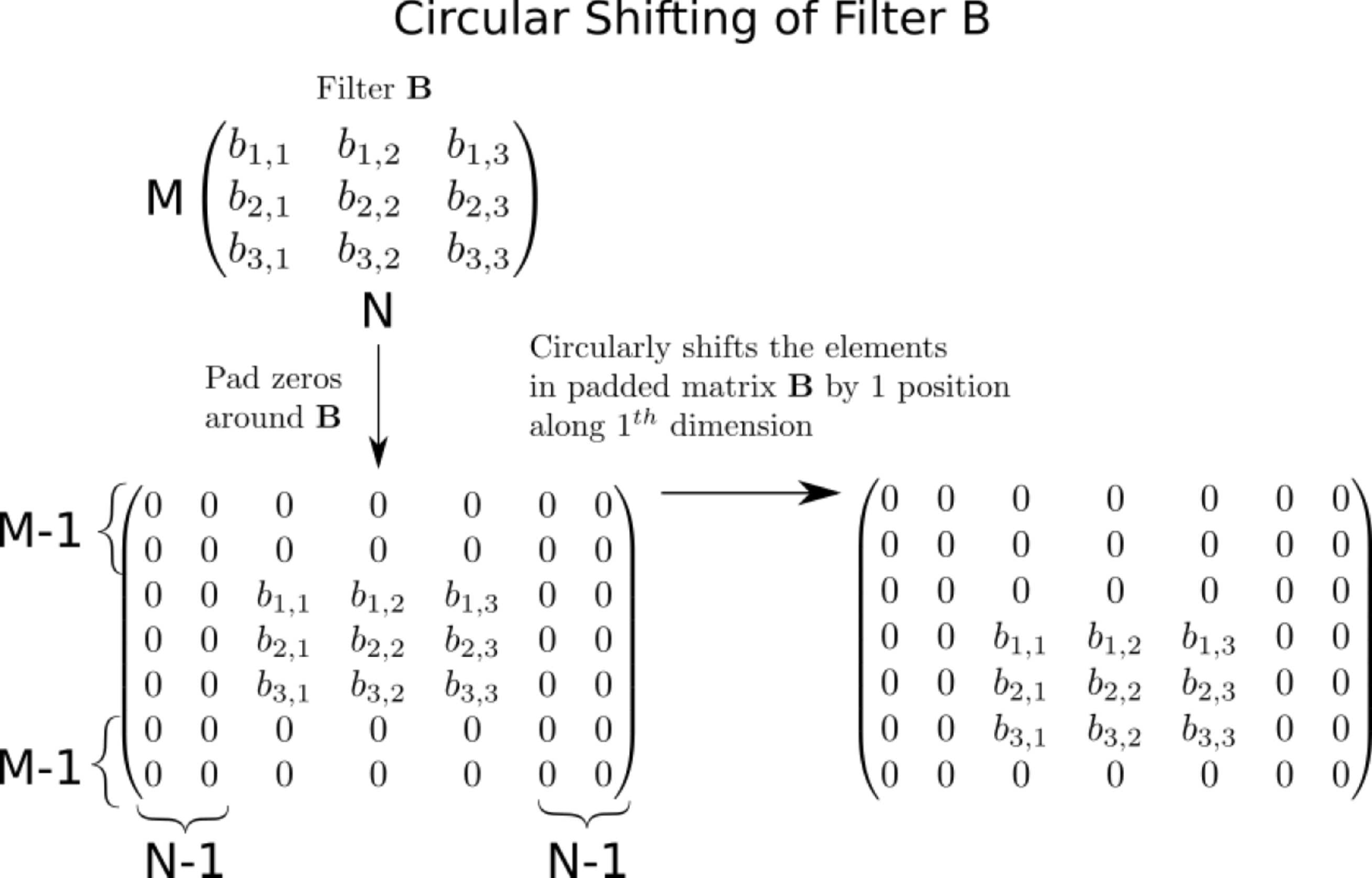}
  \caption {Diagram that illustrates the circular  shifting of padded filter $\bf B$.}
  \label{fig:circshift}
\end{figure}

To align the learned filters, we first choose  a ``target'' filter, which is denoted as a 2D matrix ${\bf A}$ with the size of $M\times N$.
Then we try to align all the other filters with the target filter through  their  cross-correlation.
For example, if we choose one filter denoted as matrix ${\bf B}$ with the same
size as ${\bf A}$, we can get the cross-correlation matrix $\bf C$ with
its elements defined as,
\begin{equation}
  { C}_{i+M,j+N}=\sum_{m=1}^{M}\sum_{j=1}^{N} { a}_{m,n}\cdot{ b}_{m+i,n+j},
  \label{eqn:xcorr2}
\end{equation}
where $-M <i< M$ and $-N< j< N$. $a_{i,j}$, $b_{i,j}$ and $C_{i,j}$ indicate the element at position $(i,j)$ in matrices
$\bf A$, $\bf B$ and $\bf C$, respectively. The location of the maximum absolute value
of the elements in matrix $\bf C$ indicates how much should we shift filter $\bf B$ to filter $\bf A$ in each direction.
Figure~\ref{fig:crossCorrelation} shows the calculation of the cross-correlation matrix $\bf C$ for two filters $A$ and $B$.
$c_{1,0}$ (or $C_{4,3}$) is the maximum absolute value of the elements in matrix $\bf C$, which indicates filter $\bf B$ should be shifted 1 position along the first direction.
Here, we need to pad zeros along all the dimensions of filter $\bf B$ before shifting it, which is displayed in
Figure~\ref{fig:circshift}.

\begin{figure}[htpb]
  \centering
  \includegraphics[width=\columnwidth]{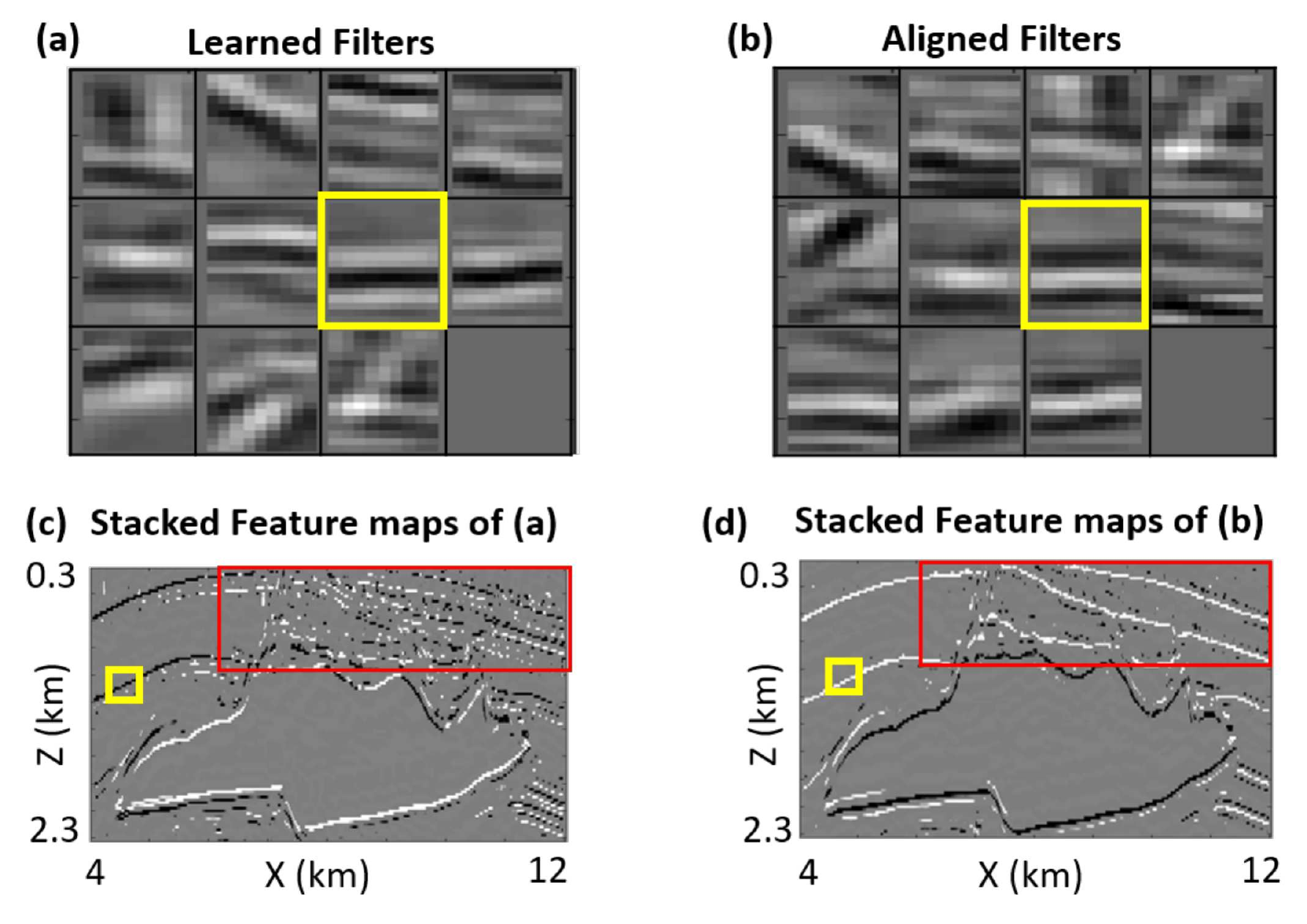}
  \caption {(a) Learned filters from the migration image of SEG/EAGE salt model; (b) the aligned filters; stacked feature maps of the (c) orginal and (d) aligned filters, where
  the yellow boxes show the sizes of the filters for the filters, respectively.}
  \label{fig:alignFilters}
\end{figure}

Figure~\ref{fig:alignFilters}a shows the learned filters with a size of $17\times9$ from the migration
image of the SEG/EAGE salt model. Filter No. 7 (yellow box) is chosen as the target filter.
The aligned filters are shown in Figure~\ref{fig:alignFilters}b without zero padding
and the stacked feature maps from the original
and aligned  filters are displayed in Figures~\ref{fig:alignFilters}c and \ref{fig:alignFilters}d,
respectively. It is evident that the reflector interfaces from the aligned filters are more continuous
 especially in the red box compared with those of the original filters.

\end{document}